\documentclass[preprint,12pt,authoryear]{elsarticle}

\makeatletter
\def\ps@pprintTitle{%
  \let\@oddhead\@empty
  \let\@evenhead\@empty
  \let\@oddfoot\@empty
  \let\@evenfoot\@oddfoot
}
\usepackage{amsthm,amssymb,amsfonts,amsmath,amstext}
\usepackage[title]{appendix}
\usepackage{bbm}
\usepackage{booktabs}
\usepackage{bookmark}
\usepackage{enumerate}
\usepackage{enumitem} 
\usepackage{graphicx}
\usepackage{indentfirst}
\usepackage{mathrsfs}
\usepackage{setspace}
\usepackage{ulem}
\usepackage{cancel}
\usepackage{xcolor}
\definecolor{mycolor}{rgb}{0.3,0.3,0.3}
\usepackage[authoryear]{natbib}
\usepackage{float}
\usepackage{pifont}
\usepackage{multirow}
\usepackage{tabularx}
\usepackage{makecell}
\usepackage{subcaption}
\usepackage{adjustbox}
\usepackage{rotating}

\usepackage{amsthm}
\newtheoremstyle{mytheorem}{4pt}{4pt}{\normalfont}{}{\bfseries}{.}{.5em}{}{\itshape}{}
\theoremstyle{mytheorem}

\usepackage{tikz}
\usetikzlibrary{arrows,backgrounds}
\tikzstyle{vertex}=[circle,draw=none,fill=black,thin,scale=0.5]
\usepackage{array}
\newcommand{\PreserveBackslash}[1]{\let\temp=\\#1\let\\=\temp}
\newcolumntype{C}[1]{>{\PreserveBackslash\centering}p{#1}}
\newcolumntype{R}[1]{>{\PreserveBackslash\raggedleft}p{#1}}
\newcolumntype{L}[1]{>{\PreserveBackslash\raggedright}p{#1}}

\providecommand{\U}[1]{\protect\rule{.1in}{.1in}}

\newtheorem{theorem}{Theorem}

\newtheorem{corollary}{Corollary}

\newtheorem{example}{Example}

\newtheorem{lemma}{Lemma}

\newtheorem{proposition}{Proposition}
\newtheorem{remark}{Remark}

\definecolor{MyDarkBlue}{rgb}{0,0.08,0.45}
\definecolor{cites}{HTML}{324b13}
\definecolor{links}{HTML}{1a663b}
\definecolor{MyLightMagenta}{cmyk}{0.1,0.8,0,0.1}
\hypersetup{colorlinks,citecolor=blue,filecolor=black,linkcolor=blue,urlcolor=blue}

\newcommand{\Xomit}[1]{}
\newcommand{\Pin}{\succ^{\mathrm{in}}}
\newcommand{\Pineq}{\succeq^{\mathrm{in}}}
\newcommand{\Pout}{\succ^{\mathrm{out}}}
\newcommand{\Pouteq}{\succeq^{\mathrm{out}}}

\newcommand{\Cin}{C^{\mathrm{in}}}
\newcommand{\Cout}{C^{\mathrm{out}}}

\begin{document}

\begin{frontmatter}

\title{Efficient Major Transition Exchange under Distributional and Dual Priority-respecting Constraints\footnote{Date: 12 April, 2025.}} 

\author[1]{Yao Cheng}

\ead{chengyao@swufe.edu.cn}

\affiliation[1]{organization={School of Economics, Southwestern University of Finance and Economics},
            city={Chengdu},
            postcode={611130}, 
            state={Sichuan},
            country={China}}

\author[2]{Jingsheng Yu}

\ead{yujingsheng@whu.edu.cn}

\affiliation[2]{organization={Economics and Management School, Wuhan University},
            city={Wuhan},
            postcode={430072}, 
            state={Hubei},
            country={China}}

\author[3]{Ling Zheng}

\ead{hyzhengling@whu.edu.cn}

\affiliation[3]{organization={Hongyi Honor College, Wuhan University},
            city={Wuhan},
            postcode={430072}, 
            state={Hubei},
            country={China}}

\begin{abstract}  
Many real matching markets encounter distributional and fairness constraints. 
Motivated by the Chinese Major Transition Program (CMT), this paper studies the design of exchange mechanisms within a fresh framework of both distributional and dual priority-respecting constraints. 
Specifically, each student has an initial assigned major and applies to transfer to a more desirable one. 
A student can successfully transfer majors only if they obtain eligibility from both their initial major and applied major. 
Each major has a dual priority: a strict priority over current students who wish to transfer out and a strict priority over students from other majors who wish to transfer in. 
Additionally, each major faces a ceiling constraint and a floor constraint to regulate student distribution. 
We show that the existing mechanism of CMT may result in inefficiencies, and propose two mechanisms that can match students to majors in an efficient way as well as respecting each major's distributional constraints and dual priority. 
The efficient mechanisms are based on a proposed solution concept: eligibility maximization (EM), and two processes for identifying improvement cycles—specifically, transfer-in exchangeable cycles and transfer-out exchangeable cycles. 

\end{abstract}


\end{frontmatter}

\section{Introduction}

Major transition is a common phenomenon worldwide because students may discover new interests or career goals, necessitating an academic shift. 
Universities and colleges regularly run major transition programs to assist students in changing their majors. 
In the Chinese Major Transition Program (CMT), each major faces a ceiling constraint and a floor constraint on the distribution of students during the transfer process, aimed at regulating a reasonable distribution of students across majors.\footnote{\ According to the data of the CMT program operated by an university in China in 2023, as shown in Table \ref{Tab_mtdata} in \ref{App_mech}, we can see that most majors face both floor and ceiling constraints.} 
The ceiling constraint restricts the number of remaining current students, and those newly transferring in, from exceeding a certain amount; 
the floor constraint requires the number of such students to remain a certain level. 
For example, the major transition policy of the Central South University of Forestry and Technology sets the floor constraint at 90\% of the program's enrollment for that year, and the ceiling constraint at 110\%.\footnote{\  Such distributional constraints can be reflected by the major transition policy of the Central South University of Forestry and Technology, which announces that the number of transfer-in students and transfer-out students should be no more than 10\% of a major's enrollment for that year, respectively; see \url{https://jwc.csuft.edu.cn/tzgg/202405/t20240515_156579.html}.}

Furthermore, majors have specific requirements for both transfer-out and transfer-in applicants. 
Those wishing to transfer out may need to meet criteria such as GPA rankings. Applicants transferring into a new major need to meet comprehensive criteria set by the prospective major, including a combination of GPA, entrance exam scores, and interview performance.\footnote{\  For instance, at Jinan University in China, students with GPAs in the top 50\% can apply for a major change, while those in the bottom 50\% need to provide additional supporting materials to bolster their applications. Moreover, each college should instruct transfer-in applicants to participate in the evaluation organized by the college. Any student who fails to attend the assessment will be regarded as a waiver of the major transition application; see \url{https://jwc.jnu.edu.cn/2024/0415/c6765a810665/page.htm}.}
To ensure fairness in the major transition program, each major typically has a dual priority: one for transfer-out applicants and another for transfer-in applicants. This dual priority plays a crucial role in determining applicants' transfer-in and transfer-out eligibility.\footnote{\ In the CMT program, once the lists of students obtaining transfer-in and transfer-out eligibility are decided, they will be publicly announced. After confirming that there are no objections, the final lists will be published.} 

Motivated by the CMT program, we introduce a framework for an exchange market under both distributional and dual priority-respecting constraints. 
Specifically, there are a set of majors and a set of students. 
Each student has an initial assigned major and applies to transfer to a more desirable one.\footnote{\ In most universities in China, students may have multiple opportunities to change majors. However, during each application process, they can specify only one preferred major. One key reason for this is that the transfer-in examinations for all majors are typically scheduled at the same time.} 
Each major has a dual priority: a strict priority over current students who wish to transfer out and a strict priority over students from other majors who wish to transfer in. 
In addition, each major faces a ceiling constraint and a floor constraint to regulate student distribution. 

An outcome can be represented by an ordered pair that contains a set of students eligible to transfer out and a set of students eligible to transfer in.  
A student can successfully transfer to their applied major only if they obtain both transfer-out eligibility and transfer-in eligibility. 
Given an outcome, there is a corresponding matching such that every student who obtains both transfer-in and transfer-out eligibility is assigned to their applied major, while all other students remain in their initial majors.  
In this paper, we focus on the set of permissible outcomes in which each major's distributional constraints and dual priority are respected.\footnote{\ We say that an outcome does not respect a major's distributional constraints if in the corresponding matching the number of students assigned to this major violates its distributional constraints. 
An outcome does not respect a major's dual priority if not every student with a higher priority than those obtaining transfer-in or transfer-out eligibility receives such eligibility in the major. 
Inspired by the existing policies of the CMT program, any outcome that does not meet the distributional and dual priority-respecting constraints is not permitted to be implemented.}

We show that the main major transition mechanism in China, referred to as the Chinese Major Transition Program with Eligibility Cap Policy (CMT-EC), may lead to inefficiencies. 
We begin by introducing the concept of Pareto efficiency, considering both distributional and dual priority-respecting constraints. 
We show that the CMT-EC mechanism may not achieve Pareto efficient outcomes because the eligibility cap policy can result in an insufficient distribution of eligibility. A motivating example is provided in Section 3 to illustrate this.

To improve the welfare of students, we aim to propose efficient major transition mechanisms which always produce permissible and Pareto efficient outcomes. We find that an efficient mechanism should provide appropriate amounts of transfer-in and transfer-out eligibility, avoiding both insufficiency and redundancy in eligibility. To address this, we introduce the concept called \textit{eligibility maximization (EM)} and show that the properties of EM provide important insights into finding Pareto efficient outcomes, despite being incompatible with Pareto efficiency. 
Specifically, given each permissible outcome which has eligibility maximization (referred to as \textit{a permissible EM outcome}), majors can be categorized into three groups: overdemanded (full transfer-out eligibility but not transfer-in), underdemanded (full transfer-in eligibility but not transfer-out), and balanced (full eligibility for both). 
We demonstrate that Pareto improvements over a permissible EM outcome can occur only through exchanges among students from either overdemanded or underdemanded majors, not across different categories. 
We further show the existence of a permissible EM outcome by proposing a major transition mechanism with eligibility maximization, which can always find a permissible EM outcome. 
 
We introduce two efficient major transition mechanisms: the TiE-prioritized efficiency-adjusted EM (EAEM-TiE) mechanism and the ToE-prioritized efficiency-adjusted EM (EAEM-ToE) mechanism.  
Specifically, starting with a permissible EM outcome, these mechanisms achieve efficiency through two types of exchange: transfer-in exchangeable cycles and transfer-out exchangeable cycles. 
We propose the process for identifying transfer-in exchangeable cycles (the TiE process) and the process for identifying transfer-out exchangeable cycles (the ToE process).  
We show that, given a permissible EM outcome as input, (1) the outcome produced by the TiE process (or the ToE process) (weakly) Pareto dominates the input; 
(2) every agent from overdemanded majors (or underdemanded majors) cannot be Pareto improved upon at the respective TiE outcome (or the respective ToE outcome). 
That is, for any permissible outcome in which Pareto dominates the TiE outcome (or the ToE outcome), the assignments for agents from overdemanded majors (or underdemanded majors) remain unchanged. 
Furthermore, we show that when starting with a permissible EM outcome, running the TiE process and the ToE process sequentially is sufficient to produce a Pareto efficient outcome. 
We also show that while the order of prioritizing the TiE process and the ToE process may result in different outcomes, both sequences lead to permissible and Pareto efficient outcomes. 
Finally, we discuss some special cases, including programs with ceiling constraints, programs with floor constraints, and programs under balanced condition.

In this paper, we provide a fresh framework for reassignment markets facing both distributional and fairness constraints,\footnote{\ The rationale for such distributional and fairness conditions varies across markets.} where agents and objects are in one organization, whether in a firm, university, or the public sector. 
In these types of market, agents seek exchanges of position, resources, etc., to improve their welfare or to comply with the organization's regulations based on their current matching.  
The framework can be applied to practical applications such as worker/teacher reassignment programs. Examples of worker/teacher reassignment programs include the Commonwealth Teacher Exchange Programme and the exchange program of the International Federation of Medical Students' Associations \citep{DU2019}, teacher reassignments in France \citep{C2022}, and job rotation in Japan \citep{YZ2020}. 
In such reassignment problems, the key challenge is to reassign agents to positions efficiently while respecting distributional and fairness concerns. 
Other potential applications include interdistrict school-choice matching programs \citep{HK2022}, daycare (nursery school) seat allocation in Japan \citep{KK2024}, and kidney exchange among multi centers \citep{Ykidney2022}. 
Notably, motivated by the major transition program in China, we apply this framework to the situation where each agent submits only one preferred major.

\textbf{Related literature.} 
Our work contributes to matching theory and is closely related to a growing literature on reassignment matching problems.\footnote{\ Matching theory has been effectively applied to a variety of allocation markets like housing market \citep{SC1974}, school choice \citep{BS1999,AS2003}, organ donation \citep{RSU2004,RSU2005,RSU2007,SUY2020}, and course allocation \citep{SU2010,EE2012}. It has also been utilized in two-sided matching markets, such as the marriage market \citep{GS1962}, medical matching \citep{R1984,RP1999}, and tuition and worker exchanges \citep{DU2019}.}  
\citet{DU2019} study two-sided exchange programs such as tuition and worker exchanges under export-import balances. 
They propose the two-sided top trading cycles mechanism and show that it is the unique mechanism that is balanced-efficient, worker-strategy-proof, acceptable, individually rational, and respecting priority bylaws regarding worker eligibility. 
Different from their work, the major transition problem is a one-sided exchange market where only students are welfare-relevant entities. Additionally, in our framework, each major has a dual priority for its current students and those wishing to transfer in, with both transfer-in and transfer-out eligibility determined endogenously. In contrast, in their work, the exchange-out eligibility is determined exogenously. 

\citet{C2022} study a teacher reassignment problem in which each teacher has an initially assigned position and seeks to exchange positions with others. 
They explore how to reassign teachers and positions via balanced exchange, and define a class of efficient and fair mechanisms, referred to as the \textit{Block Exchange} (BE) algorithm, and identify a subclass of strategy-proof mechanisms, referred to as the \textit{Teacher Optimal Block Exchange} (TO-BE) algorithm. 
\citet{KK2023} study a matching problem involving the inter-regional transfer of students under a balanced condition: for each region, the number of students transferring out to other regions equals the number of students transferring in from other regions. 
They provide a class of polynomial-time algorithms that can efficiently relocate students among individually rational and fair outcomes that respect the balanced condition. 
Fairness is a key consideration in both their work and ours.
In their models, each school has a single priority over teachers/students, whereas in our model, each major has a dual priority—one for transfer-in students and another for transfer-out students. 
Consequently, the fairness condition in our model differs from theirs. Specifically, we require fairness to be achieved within two disjointed sets of students, whereas they do not. 
Moreover, our model incorporates general distributional constraints, including both ceiling and floor constraints for majors, which encompass the exchange markets under balanced conditions. 
\citet{CTTU2025} study teacher (re)assignment problems involving both new and existing workers, considering both occupied and vacant positions available in the market. They propose a new strategy-proof mechanism that can effectively improve individual and distributional welfare over their initial matching.

This paper is also a part of the literature on matching markets under distributional constraints. 
Several studies examine matching problems under various types of ceiling constraint.
\citet{KK2015} study the design of a matching problem subject to regional ceiling constraints, where the maximum number of doctors who can be matched in each region is limited. 
Further related studies on different matching problems under ceiling constraints include \citet{KK2017a,KK2017,KK2018,Goto2014}. 
Some studies focus on matching problems under floor constraints, including \citet{biro2010college} for college admissions in Hungary, \citet{Ehlers2014} for school choice, \citet{FT2017} for the assignment of newly graduated cadets to positions in U.S. military branches, and \citet{A2021} for the Turkish public school teacher assignment market. 
More general distributional constraints are also studied \citep{Goto2017,IK2024}. 
Recently, some studies have investigated various types of distributional constraint in specific markets;   see \citet{Abraham2007} for project allocation; \citet{westkamp2013, Aygun2020} for college admission; \citet{Pathak2024} for pandemic resource allocation; \citet{delacretaz2023matching} for refugee resettlement; and \citet{delacretaz2019stability} for daycare allocation.

It is worth mentioning that our framework incorporates not only distributional constraints but also a dual priority-respecting constraint faced by each major. 
Inspired by the application of the major transition program in China, the dual priority-respecting condition, such that every student's transfer-out and transfer-in priorities are respected, is treated as a hard constraint.\footnote{\ 
Fairness is an important property in the literature of priority-based allocation problems. \textit{Elimination of justified envy}, which is a conventional fairness concept, has been widely studied in priority-based allocation problems; see \citep{BS1999,AS2003, A2020,RRS2024}. We incorporate the concept of elimination of justified envy into our dual priority system and introduce the dual priority-respecting condition (the formal definition can be found in Section 2).} 

The outline of the paper is as follows. 
Section 2 presents the model.  
Section 3 introduces the current rules of the CMT program and provides a motivating example to illustrate the inefficiency of the current rules. 
Section 4 introduces the solution concept of eligibility maximization and its properties, and a major transition mechanism with eligibility maximization. 
Section 5 introduces two efficiency-adjusted EM major transition mechanisms. 
Section 6 concludes the paper. 
Proofs are provided in the Appendix.

\section{Model}

There is a finite set of students $I$ and a finite set of majors $M$.\footnote{\ Without loss of generality, we only focus on the set of students who are interested in transferring.}  
Each student $i \in I$ has an initial major $\omega_i \in M$ and applies for a new major $\alpha_i \in M \setminus \{\omega_i\}$. It is natural to assume that every student $i$ applies to a major that can possibly accept transfers from their current major, and prefers their applied major $\alpha_i$ to their initial major $\omega_i$.
For each major $m \in M$, let $\omega_m= \{i \in I \mid \omega_i = m\}$ be the set of transfer-out applicants who are initially assigned to the major $m \in M$, and $\alpha_m = \{i \in I \mid \alpha_i = m \}$ the set of  transfer-in applicants who apply for major $m$.
Each major $m \in M$ has a strict, complete, and transitive priority $\Pout_{m}$ over its transfer-out applicants $\omega_m$, and a strict, complete, and transitive priority $\Pin_{m}$ over its transfer-in applicants $\alpha_m$. 
Let $\succ = (\Pout_m, \Pin_m)_{m \in M}$ be the priority structure. 
Every major $m \in M$ faces a ceiling constraint $\overline{q}_m$ and a floor constraint $\underline{q}_m$. 
We assume that $\underline{q}_m\leq |\omega_m|\leq \overline{q}_m$. 
We say that a major $m$ is \textit{ceiling-unbound} if $\overline{q}_m \geq |\omega_m| + |\alpha_m|$, and \textit{floor-unbound} if $\underline{q}_m = 0$. 
Let $Q = (\underline{q}_m, \overline{q}_m)_{m \in M}$ be the distributional constraint structure. 
A major transition problem is a tuple $\langle I, M, \omega, \alpha, Q, \succ \rangle$.

An outcome is \textbf{an ordered pair $(E, A)$}, where $E \subseteq I$ is the set of students who obtain the transfer-out eligibility from their initial major (referred to as \textit{the transfer-out eligibility set}), and $A \subseteq I$ is the set of students who obtain the transfer-in eligibility from their applied major (referred to as \textit{the transfer-in eligibility set}). 
Student $i \in I$ is \textbf{transferable} if $i \in E \cap A$, and \textbf{non-transferable} if $i \in I \setminus (E \cap A)$. 
For each major $m \in M$, we write $E_m = E \cap \omega_m$ and $A_m = A \cap \alpha_m$ to represent the transfer-out eligibility set and the transfer-in eligibility set of major $m$, respectively.
Let $\bar{E}_m = \omega_m \setminus E_m$ be the complement of set $E_m$ and $\bar{A}_m = \alpha_m \setminus A_m$ the complement of set $A_m$.
We also use the notations $E_{-m} = \bigcup_{m' \in M \setminus \{ m \}} E_{m'}$ and $A_{-m} = \bigcup_{m' \in M \setminus \{ m \}} A_{m'}$. 
Let $\mu {(E,A)}$ be the \textbf{corresponding assignment at outcome $(E, A)$} such that (1) every transferable student $i \in E \cap A$ has $\mu_i {(E,A)} = \alpha_i$, and (2) every non-transferable student $i \in I\setminus (E\cap A)$ has $\mu_i {(E,A)} = \omega_i$. 
We write $\mu_m (E, A)$ to represent the set of students assigned to major $m$ at $\mu {(E,A)}$. 
By definition, we have $\mu_m (E, A) = [ \omega_m \setminus (E_m \cap A)] \cup [A_m \cap E] $.

An outcome $(E, A)$ \textbf{respects major $m$'s distributional constraints} 
if the corresponding assignment $\mu {(E, A)}$ at $(E, A)$ satisfies $\underline{q}_m \leq \vert \mu_m (E, A) \vert \leq \overline{q}_m$. 

Given an outcome $(E, A)$, major $m$'s transfer-out priority $\Pout_m$ is violated if there exist two students $i, i' \in \omega_m$ such that $i \Pout_m i'$, $i \notin E$, and $i' \in E$; major $m$'s transfer-in priority $\Pin_m$ is violated if there exist two students $i, i' \in \alpha_m$ such that $i \Pin_m i'$, $i \notin A$, and $i' \in A$. 
In other words, a violation of the transfer-out priority (the transfer-in priority, respectively) of major $m$ at outcome $(E, A)$ occurs when there is an agent with a higher priority than an agent who obtains transfer-out eligibility (or transfer-in eligibility), but this higher-priority agent does not receive transfer-out eligibility (transfer-in eligibility, respectively) at $(E, A)$. 
We say that an outcome $(E, A)$ \textbf{respects major $m$'s dual priority} if both the transfer-out priority and the transfer-in priority of major $m$ are \textit{not} violated.

An outcome $(E, A)$ is \textbf{permissible} if it respects $m$'s distributional constraints and dual priority for each major $m \in M$. 
A permissible outcome not only considers the distributional constraints of each major, but also fairness concerns, by requiring respect for the dual priority faced by each major. 
Motivated by the CMT program, where fairness in the transition process, together with distributional constraints, are hard constraints, this paper focuses on the set of permissible outcomes.

A permissible outcome $(E,A)$ is Pareto dominated by another permissible outcome $(E',A')$ if $(E \cap A) \subsetneq (E' \cap A')$. A permissible outcome is \textbf{(constrained) Pareto efficient} if it cannot be Pareto dominated by any other permissible outcome.

A major transition mechanism $\varphi$ produces a \textit{permissible} outcome for every problem. 
A major transition mechanism $\varphi$ is \textbf{efficient} if, for every problem, it produces a permissible and Pareto efficient outcome.

\section{Motivations}

In this section, we demonstrate that the Chinese Major Transition Program with Eligibility Cap Policy (CMT-EC), which is a main rule applied in the CMT program, may lead to inefficiencies.

\subsection{Rules of the CMT-EC mechanism}

In the CMT-EC mechanism, the procedure for each major to decide the transfer-out eligibility set and the transfer-in eligibility set is as follows: 

\begin{itemize}

    \item Eligibility Cap Announcement: Each major $m$ announces a fixed transfer-out eligibility cap $p^o_m$ to limit the size of the transfer-out eligibility set, and a fixed transfer-in eligibility cap $p^i_m$ to limit the size of the transfer-in eligibility set.

    \item Transfer-out Eligibility:
    Students who are interested in transferring can apply for transfer-out eligibility.  
    Each major $m$ assigns transfer-out eligibility to at most the top $p_m^o$ transfer-out applicants based on its transfer-out priority.\footnote{\ The transfer-out priority of each major is normally dicated by criteria such as the student's GPA, overall academic performance, and specific experiences (e.g., military service).}

    \item Transfer-in Eligibility: 
    Each student who obtains transfer-out eligibility applies to one preferred major that can possibly accept transfers from their current major. 
    Each major $m$ assigns transfer-in eligibility to at most the top $p_i^o$ applicants based on its transfer-in priority.\footnote{\ Each major normally conducts an assessment, often through exams, to establish a transfer-in priority over transfer-in applicants.} 

    \item Result: The transfer-out eligibility set and the transfer-in eligibility set will be publicly announced.
After confirming that there are no objections, applicants who obtain both transfer-out eligibility and transfer-in eligibility can make their transfer successfully. Any other applicants stay put.

\end{itemize}

The primary purpose of the Eligibility Cap (EC) policy is to regulate a reasonable distribution of students across majors. Inspired by this purpose, we translate each major $m$'s announced transfer-out eligibility cap $p_m^o$ and transfer-in eligibility cap $p_m^i$ into $m$'s distributional constraints. 
For each major $m$, we define the floor and ceiling constraints under the announced caps $p^o_m$ and $p^i_m$ as follows: 
$$
\underline{q}_m = \max\{0, \vert \omega_m\vert - p^o_m\}, \quad
\overline{q}_m = |\omega_m| + p^i_m.
$$

The floor constraint $\underline{q}_m$ represents the minimum possible number of students remaining in major $m$ after transfers, under the announced caps, assuming that no student transfers in and $\min\{\vert \omega_m \vert, p_m^o\}$ number of transfer-out applicants transfer out.
The ceiling constraint $\overline{q}_m$ represents the maximum possible number of students remaining in major $m$ after transfers, under the announced caps, assuming that $q_m^i$ number of transfer-in applicants transfer in and no student transfers out. 

For example, suppose that a major $m$ has two current students $i_1, i_2$ who want to transfer out, and three students $i_3, i_4, i_5$ who apply to transfer in. Major $m$ announces $(p_m^o, p_m^i)=(1,1)$. 
Therefore, $m$'s floor constraint is $\underline{q}_m = \max\{0, \vert \omega_m \vert - p^o_m \}=\max\{0,2-1\}=1$, and the ceiling constraint is $\overline{q}_m = \vert \omega_m\vert + p^i_m =2+1=3$. 
This means that after the transfers, the number of students in major $m$ must lie within the interval $[1,3]$. 
If student $i_1$ transfers out and students $i_3$ and $i_4$ transfer in, then the number of students in major $m$ after the transfers is $2 - 1 + 2 = 3$, which satisfies the distributional constraints. 
However, if student $i_1$ transfers out and students $i_3$, $i_4$, and $i_5$ all transfer in, then the number of students becomes $2 - 1 + 3 = 4$, which violates the ceiling constraint. 

In the CMT-EC mechanism, the transfer-out eligibility set and the transfer-in eligibility set will be publicly announced.
After confirming that there are no objections, the final outcome will be published. 
Therefore, the outcome produced by the CMT-EC mechanism (referred to as \textit{the CMT-EC outcome}) respects each major's distributional constraints and dual priority; that is, the CMT-EC outcome is a permissible outcome.  


\subsection{A Motivating Example}

In the CMT-EC mechanism, since the eligibility caps announced by each major are fixed, the number of students who can obtain transfer-out eligibility and transfer-in eligibility from major $m$ are $\min \{ q_m^o, \vert \omega_m \vert \}$ and $\min \{ q_m^i, \vert \alpha_m \vert \}$, respectively. 
However, 
the fixed eligibility cap policy may result in insufficient eligibility distribution on both the transfer-in and the transfer-out sides, leading to inadequate major exchanges among students. 
We use the following example to illustrate this. 

\begin{example}\label{Ex_cmp01}
    Let $M=\{m_1,m_2,m_3\}$ be a set of majors. 
    Suppose that in the stage of eligibility cap announcement, the transfer-in and transfer-out eligibility caps announced by each major $m \in \{m_1, m_2, m_3\}$ are $(p^o_m, p^i_m)=(1, 1)$.  
    Suppose that students $i_1, i_2, i_3, i_4$ wish to transfer majors, where $\omega_{i_1}=\omega_{i_2}=m_1$, $\omega_{i_3}=m_2$, $\omega_{i_4}=m_3$, $\alpha_{i_3}=\alpha_{i_4}=m_1$, $\alpha_{i_1}=m_2$, and $\alpha_{i_2}=m_3$. 
    By definitions, the distributional constraints faced by each major are $(\underline{q}_{m_1}, \overline{q}_{m_1})=(1, 3)$ and $(\underline{q}_{m_2}, \overline{q}_{m_2})=(\underline{q}_{m_3}, \overline{q}_{m_3})=(0, 2)$.

In the stage of determining transfer-out eligibility, students $i_1, i_2, i_3, i_4$ apply for transfer-out eligibility from their current major. 
Suppose that each major $m$'s transfer-out priority $\Pout_m$ is as follows:  
\[
\Pout_{m_1}: i_1, i_2 \qquad \Pout_{m_2}: i_3 \qquad \Pout_{m_3}: i_4.
\] 
According to the announced eligibility caps, each major distributes at most one eligibility to transfer-out applicants. Therefore, major $m_1$ distributes transfer-out eligibility to student $i_1$; major $m_2$ distributes transfer-out eligibility to student $i_3$; major $m_3$ distributes transfer-out eligibility to student $i_4$. 

In the stage of determining transfer-in eligibility, students $i_1, i_3, i_4$, who obtain transfer-out eligibility, apply to their desirable major. That is, $i_1$ applies to $m_2$; $i_3$ applies to $m_1$; $i_4$ applies to $m_1$. 
Suppose that each major $m$'s transfer-in priority $\Pin_m$ is as follows:  
\[
\Pin_{m_1}: i_3, i_4 \qquad \Pin_{m_2}: i_1.
\]

According to the announced eligibility caps, each major distributes at most one eligibility to transfer-in applicants. 
Therefore, major $m_1$ distributes transfer-in eligibility to student $i_3$; 
major $m_2$ distributes transfer-in eligibility to student $i_1$. 

In summary, the CMT-EC outcome is $(E, A)=(\{i_1, i_3, i_4\}, \{i_1, i_3\})$ and the corresponding assignment is 
$\mu(E, A) = \left( \begin{smallmatrix}     i_1 & i_2 & i_3 & i_4\\ m_2 & m_1 & m_1 & m_3 \end{smallmatrix} \right)$. 
\end{example}

The CMT-EC mechanism may result in insufficient eligibility on both the transfer-in side and the transfer-out side. 
On the one hand, the transfer-in eligibility cap may result in insufficient eligibility on the transfer-in side. 
In this example, student $i_4$ obtains transfer-out eligibility from major $m_3$ but fails to obtain transfer-in eligibility from major $m_1$ because each major gives at most one student their transfer-in eligibility according to their announced eligibility caps. 
However, major $m_1$ distributing student $i_4$ transfer-in eligibility does not violate its distributional constraints. 

On the other hand, the transfer-out eligibility cap may lead to insufficient eligibility on the transfer-out side, and the rules stipulating that only transfer-out eligible students can apply for their desired majors also lead to insufficient eligibility on the transfer-in side. 
In this example, since student $i_2$ does not obtain transfer-out eligibility and hence does not make any application. Suppose that $i_2$ can make an application to $m_3$ and the transfer-in priority of each major is 
\[
\Pin_{m_1}: i_3, i_4 \qquad \Pin_{m_2}: i_1 \qquad \Pin_{m_3}: i_2 
\] 
Then, major $m_1$ giving student $i_2$ transfer-out eligibility and major $m_3$ giving $i_2$ transfer-in eligibility does not violate any major's distributional and dual priority-respecting constraints. 
In this case, the CMT-EC outcome $(E, A)$ can be Pareto dominated by the Pareto efficient outcome $(E^*, A^*)=(\{i_1, i_2, i_3, i_4\}, \{i_1, i_2, i_3, i_4\})$, where the corresponding assignment is $\mu(E^*, A^*)= \left( \begin{smallmatrix} i_1 & i_2 & i_3 & i_4\\ m_2 & m_3 & m_1 & m_1 \end{smallmatrix} \right) $.

Inspired by these observations, this paper aims to design an efficient mechanism, subject to each major's distributional and dual priority-respecting constraints.
To tackle insufficient eligibility distribution, an intuitive idea is that each major maximizes the distribution of transfer-in eligibility and transfer-out eligibility as much as possible as long as its distributional and dual priority constraints are respected. 
However, we find that such a solution may lead to redundant eligibility, causing some students to become interrupters and preventing the achievement of a Pareto efficient outcome (as analyzed in detail in Section \ref{Section_EM}).

In summary, an efficient mechanism should ensure neither insufficient eligibility distribution, which could hinder full student exchange, nor redundant eligibility, which could make some students obstruct others' transfer opportunities. 
In the next two sections, we will show how to design an efficient major transition mechanism.

\section{Eligibility Maximization}\label{Section_EM}

In this section, we introduce the concept of \textit{eligibility maximization (EM)} and show the favorable properties of a permissible EM outcome. 
We emphasize that although eligibility maximization does not imply Pareto efficiency, it offers a valuable insight into finding a Pareto efficient outcome. 
Therefore, before introducing an efficient mechanism, we first introduce the EM concept and show the existence of permissible EM outcomes by proposing a major transition mechanism with eligibility maximization. 

\subsection{Eligibility Maximization and its Properties}

As discussed above, an intuitive idea to maximize students' eligibility is that each major expands its transfer-out eligibility set and transfer-in eligibility set as long as their dual priority-respecting and distributional constraints are respected. 
We say that a major $m$ can \textbf{permissibly expand} its transfer-out eligibility set and transfer-in eligibility set at an outcome $(E, A)$ if there exists $E'_m \supseteq E_m$ and $A'_m \supseteq A_m$ with at least one proper superset, such that the new outcome $(E',A') = (E'_m \cup E_{-m}, A'_m \cup A_{-m} )$ respects $m$'s dual priority and distributional constraints.\footnote{\ Such expansion can be regarded as a decentralized process, because each major only focuses on their own dual priority-respecting and distributional constraints when distributing their transfer-in and transfer-out eligibility.} 
An outcome has \textbf{eligibility maximization (EM)} when no major has such a permissible expansion. 
In an EM outcome, if a major expands either its transfer-out eligibility set or transfer-in eligibility set, or both, it will inevitably violate either its dual priority-respecting constraint or its distributional constraints. 
We say that a major transition mechanism $\varphi$ has \textbf{eligibility maximization} if for every problem, it produces a permissible EM outcome. 

Notably, permissibility and eligibility maximization do not imply each other. 
We use an example to illustrate this.

\begin{example}\label{Ex_EMP}(\textit{Eligibility maximization $\nRightarrow$ permissibility, and vice versa}) 
    There are two majors $M = \{m_1,m_2\}$ with distributional constraints $(\underline{q}_{m_1}, \overline{q}_{m_1})=(1,1)$ and $(\underline{q}_{m_2}, \overline{q}_{m_2})=(0,1)$, and one student $i$ who has $\omega_i = m_1$ and $\alpha_i = m_2$. 
    Consider a permissible outcome $(E, A)=(\{i\},\emptyset)$. 
    Major $m_2$ can permissibly expand its transfer-out eligibility set and transfer-in eligibility set via the outcome $(E', A')=(\{i\},\{i\})$ since the new outcome $(E', A')$ respects the dual priority and distributional constraints of $m_2$. 
    This indicates that a permissible outcome does not imply an EM outcome. 
    Moreover, outcome $(E', A')$ is an EM outcome but not a permissible outcome because the floor constraint of $m_1$ is violated. 
    In this case, a permissible EM outcome is $(E'', A'') = (\emptyset, \{i\})$.
\end{example}

Fortunately, for every major transition problem, the intersection of permissible outcomes and EM outcomes is non-empty.  
Formally, 

\begin{proposition}\label{Prop_EMP}
For every problem $\langle I, M, \omega, \alpha, Q, \succ \rangle$, there exists at least one permissible EM outcome. 
\end{proposition}

However, Pareto efficiency and eligibility maximization are incompatible. 

\begin{proposition}\label{prop_incom}
There may not exist a permissible outcome that is Pareto efficient and has eligibility maximization. 
\end{proposition}
\proof
We use the following example to prove this. 
\begin{example}\label{Ex_incom}
There are five majors $M=\{m_1, m_2, m_3,m_4, m_5\}$ and seven students $I=\{i_1, i_2,..,i_7\}$. The distributional constraints, the transfer-out priority, and the transfer-in priority of each major are given in the following table.
\begin{center}
\begin{tabular}{|c|c|l|l|}
\hline
    Major & $(\underline{q}_m, \overline{q}_m)$ & transfer-out priority & transfer-in priority \\\hline
    $m_1$ & $(0,1)$  & $\Pout_{m_1}: i_1$ & $\Pin_{m_1}: i_3, i_2$ \\\hline
    $m_2$ & $(0,1)$  & $\Pout_{m_2}: i_2$ & $\Pin_{m_2}: i_7, i_1$ \\\hline
    $m_3$ & $(2,2)$  & $\Pout_{m_3}: i_3, i_4 $  & $\Pin_{m_3}: i_6$ \\\hline
    $m_4$ & $(2,2)$  & $\Pout_{m_4}: i_5, i_6$  & $\Pin_{m_4}: i_4$ \\\hline
    $m_5$ & $(0,2)$  & $\Pout_{m_5}: i_7 $  & $\Pin_{m_5}: i_5$ \\\hline
\end{tabular}
\end{center} 
\end{example}

There are two permissible and Pareto efficient outcomes. 
We will show that the two outcomes do not have eligibility maximization. 
The first permissible and Pareto efficient outcome is $(E^*, A^*)=(\{i_1, i_2\}, I)$. 
The outcome $(E^*, A^*)$ does not have eligibility maximization because major $m_5$ can permissibly expand its transfer-out eligibility set and transfer-in eligibility set via $(E'_{m_5}, A'_{m_5})$, where $E'_{m_5}=\{i_7\} \supset E^*_{m_5}=\emptyset$ and $A'_{m_5}=A^*_{m_5}$, such that the new outcome $(E',A') = (E'_{m_5} \cup E^*_{-m_5}, A^*)$ respects $m_5$'s dual priority and distributional constraints. 
The second permissible and Pareto efficient outcome is $(E^\diamond, A^\diamond)=(I, \{i_4, i_6\})$. 
The outcome $(E^\diamond, A^\diamond)$ does not have eligibility maximization because major $m_5$ can permissibly expand its transfer-out eligibility set and transfer-in eligibility set via $(E'_{m_5}, A'_{m_5})$, where $E'_{m_5}= E^\diamond_{m_5}$ and $A'_{m_5}=\{i_5\} \supset E^\diamond_{m_5}=\emptyset$, such that the new outcome $(E',A') = (E^\diamond, A'_{m_5} \cup A^\diamond_{-m_5})$ respects $m_5$'s dual priority and distributional constraints. 
\qed

Although, by Proposition \ref{prop_incom}, a permissible EM outcome may not be Pareto efficient, the following lemma and proposition offer insights into designing efficient mechanisms based on a permissible EM outcome. 
Specifically, Lemma~\ref{Le_AT} demonstrates that in a permissible EM outcome, majors can be categorized into three groups --- overdemanded majors, underdemanded majors, and balanced majors --- and Proposition~\ref{Prop_AT2} outlines how agents in each group can achieve Pareto improvements beyond a permissible EM outcome. 
Formally,

\begin{lemma}\label{Le_AT}
    Given a permissible EM outcome $(E, A)$, each major $m \in M$ satisfies at least one of the following: $E_m = \omega_m$ or $A_m = \alpha_m$.
\end{lemma}

As a consequence of Lemma~\ref{Le_AT}, in a permissible EM outcome $(E,A)$, the set of majors can be divided into the following three disjoint subsets:
\begin{align*}
    M^{O} (E, A) & = \{m \in M \mid E_m = \omega_m \mbox{ and } A_m \neq \alpha_m \}, \\
    M^{U} (E, A) & = \{m \in M \mid E_m \neq \omega_m \mbox{ and } A_m = \alpha_m\}, \\
    M^{F} (E, A) & = \{m \in M \mid E_m = \omega_m \mbox{ and } A_m = \alpha_m \}.
\end{align*}

Each major $m \in M^O(E, A)$ is \textit{overdemanded} at $(E, A)$, such that every student applying to transfer out of $m$ receives transfer-out eligibility, but not every student applying to transfer into $m$ receives transfer-in eligibility. This implies that major $m$ must have $\vert \mu_m (E, A) \vert = \overline{q}_m$. 
Otherwise, major $m$ can additionally distribute transfer-in eligibility to the highest priority agent $i=\max_{\Pin_m} \bar{A}_m$ without violating $m$'s dual priority and distributional constraints, which causes the contradiction that $(E, A)$ is a permissible EM outcome. 
Let $I^O (E,A) = \{i \in I\mid \mu_i (E,A) \in M^O (E,A)\}$ be the set of students assigned to overdemanded majors at $(E, A)$.

Each major $m \in M^U(E, A)$ is \textit{underdemanded} at $(E, A)$, such that every student applying to transfer into $m$ receives transfer-in eligibility, but not every student applying to transfer out of $m$ obtains transfer-out eligibility. 
This implies that major $m$ must have $\vert \mu_m (E, A) \vert = \underline{q}_m$. Otherwise, major $m$ can additionally distribute transfer-out eligibility to the highest priority agent $i=\max_{\Pout_m} \bar{E}_m$ without violating $m$'s dual priority and distributional constraints, which causes the contradiction that $(E, A)$ is a permissible EM outcome. 
Let $I^U (E,A) = \{i \in I\mid \mu_i (E,A) \in M^U (E,A) \}$ be the set of students assigned to underdemanded majors at $(E, A)$. 

Each major $m \in M^F(E, A)$ is \textit{balanced} at $(E, A)$, such that every student applying to transfer out of $m$ receives transfer-out eligibility, and every student applying to transfer into $m$ receives transfer-in eligibility. 
Let $I^F (E,A) = \{i \in I\mid \mu_i (E,A) \in M^F (E,A) \}$ be the set of students assigned to balanced majors at $(E, A)$.

The following proposition provides an important perspective on how to achieve Pareto improvements based on a permissible EM outcome.

\begin{proposition}\label{Prop_AT2}
    If a permissible outcome $(E',A')$ Pareto dominates a permissible EM outcome $(E,A)$, then the following statements hold true: \\    
    (1) For each agent $i \in I$, $\mu_i {(E, A)} \in M^O(E, A)$ if and only if $\mu_i {(E', A')} \in M^O(E, A)$; \\
    (2) For each agent $i \in I$, $\mu_i {(E, A)} \in M^U(E, A)$ if and only if $\mu_i {(E', A')} \in M^U(E, A)$; \\
    (3) For each agent $i \in I$, $\mu_i {(E,A)} \in M^F(E, A)$ implies $\mu_i {(E',A')} = \mu_i {(E,A)}$; \\   
    (4) For each major $m \in M$, $\vert \mu_m (E, A) \vert = \vert \mu_m (E', A') \vert$ holds.
\end{proposition}

For any outcome $(E', A')$ that Pareto dominates a permissible EM outcome $(E, A)$, agents from $I^O(E, A)$ (or $I^U(E, A)$) can achieve a Pareto improvement only by swapping majors among themselves. In contrast, the agents in $I^F(E, A)$ do not achieve any Pareto improvement. Moreover, the statement (4) implies that Pareto improvements occur through cycles. Specifically, if a student $i_0 \notin E \cap A$ successfully transfers to major $\alpha_{i_0}$ at $(E', A')$, then there must be a student $i_1 \in \omega_{\alpha_{i_0}}$ with $i_1 \notin E \cap A$ who also transfers successfully at $(E', A')$. 
Continuing in this way, this process eventually reaches an agent $i_m \notin E \cap A$ whose transfer closes the cycle, with $\alpha_{i_m}=\omega_{i_0}$.

\subsection{A Major Transition Mechanism with Eligibility Maximization}

By Proposition \ref{Prop_AT2}, an intuitive idea to design an efficient mechanism is to start with a permissible EM outcome and seek further Pareto improvements among students through cycles. 
By Proposition \ref{Prop_EMP}, we know that a permissible EM outcome always exists, and hence a primary question is: how can we find a permissible EM outcome? 
In this section, we introduce a major transition mechanism with eligibility maximization (referred to as \textit{the EM mechanism}) that produces a permissible EM outcome for every problem.\footnote{\ Mechanisms with eligibility maximization are not unique, and an alternative mechanism is provided in the~\ref{App_mech}.} 

We say that a major $m$ is \textbf{transfer-in expandable} at an outcome $(E, A)$ if (1) $\bar{A}_m \neq \emptyset$;
(2) either $\max_{\Pin_m} \bar{A}_m \notin E$ or $\vert \mu_m (E, A) \vert < \overline{q}_m$. 
Let $M^{\mathrm{in+}}(E, A)$ be the set of transfer-in expandable majors at outcome $(E, A)$. 



The EM mechanism starts with a permissible outcome in which every student has transfer-out eligibility but no transfer-in eligibility. 
It reaches a permissible EM outcome through two iterative processes: the transfer-in process and the transfer-out process. 
Since each major initially gives transfer-out eligibility to all their transfer-out applicants, the focus is on finding transfer-in expandable majors. During the transfer-in process, each transfer-in expandable major distributes transfer-in eligibility to the highest-priority transfer-in applicants who have not yet received it. 
In the transfer-out process, if a major's floor constraint is violated after transfer-in adjustments, the major resolves this by revoking transfer-out eligibility from the lowest-priority transfer-out applicants who currently have transfer-out eligibility (possibly multiple times), thereby satisfying the floor constraint. 
This adjustment may create vacancies, making some majors transfer-in expandable again, and the transfer-in process is then repeated. 
The algorithm continues iteratively until it reaches equilibrium, where no major is transfer-in expandable, and no major's floor constraint is violated. 
If transfer-in expandable majors and majors whose floor constraints are violated exist simultaneously, the algorithm prioritizes the transfer-in process. 
Let $M^{\mathrm{out-}}(E, A)=\{m \in M \mid \vert \mu_m(E, A) \vert < \underline{q}_m \}$ be the set of majors whose floor constraint is violated at $(E, A)$. 
A formal description of the EM mechanism is as follows.

\begin{center}
    \textbf{A major transition mechanism with eligibility maximization}
\end{center}
\begin{itemize}    
     \item[\textbf{Step $0$}] Let $E^1 = I$, $A^1 = \emptyset$. 
    
    \item[\textbf{Step $k$}]($k \geq 1$) Consider the outcome $(E^k, A^k)$. 
    Do the following processes:  
    \begin{itemize}
        \item[$k.1$] (\textit{Transfer-in process})
        If $M^{\mathrm{in+}}(E^k, A^k) = \emptyset$, go to Step $k.2$. 
        Otherwise, each major $m \in M^{\mathrm{in+}}(E^k, A^k)$ distributes the transfer-in eligibility to the student $\max_{\Pin_m} \bar{A}^k_m$ who has the highest priority among those transfer-in applicants who have not yet obtained transfer-in eligibility from $m$. 
        Set $A^{k+1} = A^k \cup  (\max_{\Pin_m} \bar{A}^k_m)_{m \in M^{\mathrm{in+}}(E^k, A^k)}$ and $E^{k+1} = E^k$. Let $k=k+1$, and go to step $k$. 
    
        \item[$k.2$] (\textit{Transfer-out process})
        If $M^{\mathrm{out-}}(E^k, A^k)=\emptyset$, stop.  
        Otherwise, each major $m \in M^{\mathrm{out-}}(E^k, A^k)$ revokes the transfer-out eligibility of the student $\min_{\Pout_m} E^k_m$ who has the lowest priority among those transfer-out applicants who have obtained transfer-out eligibility from $m$. 
        Set $E^{k+1} = E^{k} \setminus (\min_{\Pout_m} E^k_m)_{m \in M^{\mathrm{out-}}(E^k, A^k)}$ and $A^{k+1}=A^{k}$. 
        Let $k = k+1$, and go to step $k$. 
    \end{itemize}    
\end{itemize}

According to the processes of the EM mechanism, we have the following remark. 

\begin{remark}\label{Remark_EM}
Let $K$ be the last step of the EM mechanism. That $E^{k+1} \subseteq E^{k}$ and $A^{k} \subseteq A^{k+1}$ hold true for each step $1 \leq  k <  K$.   
\end{remark} 

In other words, during the EM mechanism, the transfer-out eligibility set is (weakly) monotonically decreasing, and the transfer-in eligibility set is (weakly) monotonically increasing. 

Moreover, according to the EM mechanism, the outcome produced at each step does not violate any major's dual priority, and the final outcome is obtained when no major is transfer-in expandable and no major's floor constraint is violated. Therefore, the following proposition holds. 

\begin{proposition}\label{Prop_M-EM1}
The EM mechanism produces a permissible EM outcome for every problem.  
\end{proposition}

We use Example \ref{Ex_incom} to show how the EM mechanism works. 
Table~\ref{Tab_Ex4.4} shows the corresponding processes of the EM mechanism.\footnote{\ In the second column of Table~\ref{Tab_Ex4.4}, the upper part represents the transfer-out eligibility set $E^k$ at each step $k$, and the lower part represents the transfer-in eligibility set $A^k$ at each step $k$.
For the upper part ($E^k$), a black mark indicates that the agent has obtained transfer-out eligibility, while a cross indicates that the agent's transfer-out eligibility has been revoked.
For the lower part ($A^k$), a black mark indicates that the agent has obtained transfer-in eligibility, whereas a grey mark indicates that the agent has not obtained transfer-in eligibility.}

\vspace{0.2cm}
\begin{table}[ht]\small
\setlength{\abovecaptionskip}{0cm}
\setlength{\belowcaptionskip}{-0.2cm} 
\caption{\small{The process of the EM mechanism in Example \ref{Ex_incom}}}
\vspace{0.2cm}
\label{Tab_Ex4.4}
\begin{tabular}{|c|c|c|c|c|}
\hline
    Step & $E^k \& A^k$ & $\mu (E^{k}, A^{k})$ & $M^{\mathrm{in+}}(E^k, A^k)$ & $M^{\mathrm{out-}}(E^k, A^k)$\\\hline
   
    $1$ & 
    \makecell[c]{ 
    $\big\{ \{i_1\}, \{i_2\}, \{i_3, i_4\}, \{i_5,i_6\}, \{i_7\} \big\}$ \\ 
    $\big\{ \{ \textcolor{mycolor}{i_3, i_2} \}, \{ \textcolor{mycolor}{i_7, i_1} \}, \{ \textcolor{mycolor}{i_6} \}, \{ \textcolor{mycolor}{i_4} \}, \{ \textcolor{mycolor}{i_5} \} \big\}$
    } & \scalebox{1.3}{$\left( \begin{smallmatrix} m_1 & m_2 & m_3 & m_4 & m_5\\ i_1 & i_2 & i_3, i_4 & i_5, i_6 & i_7 \end{smallmatrix}  \right)$} & $\{m_5\}$& $ \emptyset $ \\\hline

    $2$ & 
    \makecell[c]{ 
    $\big\{ \{i_1\}, \{i_2\}, \{i_3, i_4\}, \{i_5,i_6\}, \{i_7\} \big\}$ \\ $\big\{ \{ \textcolor{mycolor}{i_3, i_2} \}, \{ \textcolor{mycolor}{i_7, i_1} \}, \{ \textcolor{mycolor}{i_6} \}, \{ \textcolor{mycolor}{i_4} \}, \{ i_5\textcolor{mycolor}{} \} \big\}$ }& 
    \scalebox{1.3}{$\left( \begin{smallmatrix} m_1 & m_2 & m_3 & m_4 & m_5\\ i_1 & i_2 & i_3, i_4 & i_6 & i_7, i_5 \end{smallmatrix}  \right)$} & $\{m_4\}$& $ \{m_4\} $ \\\hline

    $3$ & 
    \makecell[c]{ 
    $\big\{ \{i_1\}, \{i_2\}, \{i_3, i_4\}, \{i_5,i_6\}, \{i_7\} \big\}$ \\ $\big\{ \{ \textcolor{mycolor}{i_3, i_2} \}, \{ \textcolor{mycolor}{i_7, i_1} \}, \{ \textcolor{mycolor}{i_6} \}, \{ i_4\textcolor{mycolor}{} \}, \{ i_5\textcolor{mycolor}{} \} \big\}$} &  
    \scalebox{1.3}{$\left( \begin{smallmatrix} m_1 & m_2 & m_3 & m_4 & m_5\\ i_1 & i_2 & i_3  & i_6,i_4 & i_7, i_5 \end{smallmatrix}  \right)$} & $\{m_3\}$& $ \{m_3\} $ \\\hline

    $4$ & 
    \makecell[c]{ 
    $\big\{ \{i_1\}, \{i_2\}, \{i_3, i_4\}, \{i_5,i_6\}, \{i_7\} \big\}$ \\ $\big\{ \{ \textcolor{mycolor}{i_3, i_2} \}, \{ \textcolor{mycolor}{i_7, i_1} \}, \{ i_6\textcolor{mycolor}{} \}, \{ i_4\textcolor{mycolor}{} \}, \{ i_5\textcolor{mycolor}{} \} \big\}$ }&  
    \scalebox{1.3}{$\left( \begin{smallmatrix} m_1 & m_2 & m_3 & m_4 & m_5\\ i_1 & i_2 & i_3,i_6  & i_4 & i_7, i_5 \end{smallmatrix}  \right)$} & $\emptyset$& $\{m_4\}$ \\\hline

    $5$ & 
    \makecell[c]{ 
    $\big\{ \{i_1\}, \{i_2\}, \{i_3, i_4\}, \{i_5,\xcancel{i_6}\}, \{i_7\} \big\}$ \\ $\big\{ \{ \textcolor{mycolor}{i_3, i_2} \}, \{ \textcolor{mycolor}{i_7, i_1} \}, \{ i_6\textcolor{mycolor}{} \}, \{ i_4\textcolor{mycolor}{} \}, \{ i_5\textcolor{mycolor}{} \} \big\}$ }& 
    \scalebox{1.3}{$\left( \begin{smallmatrix} m_1 & m_2 & m_3 & m_4 & m_5\\ i_1 & i_2 & i_3  & i_6,i_4 & i_7, i_5 \end{smallmatrix}  \right)$} & $\emptyset$& $\{m_3\}$ \\\hline

    $6$ & 
    \makecell[c]{ 
    $\big\{ \{i_1\}, \{i_2\}, \{i_3, \xcancel{i_4}\}, \{i_5,\xcancel{i_6}\}, \{i_7\} \big\}$ \\ $\big\{ \{ \textcolor{mycolor}{i_3, i_2} \}, \{ \textcolor{mycolor}{i_7, i_1} \}, \{ i_6\textcolor{mycolor}{} \}, \{ i_4\textcolor{mycolor}{} \}, \{ i_5\textcolor{mycolor}{} \} \big\}$}& 
    \scalebox{1.3}{$\left( \begin{smallmatrix} m_1 & m_2 & m_3 & m_4 & m_5\\ i_1 & i_2 & i_3,i_4  & i_6 & i_7, i_5 \end{smallmatrix}  \right)$} & $\emptyset$& $\{m_4\}$ \\\hline

    $7$ & 
    \makecell[c]{ 
    $\big\{ \{i_1\}, \{i_2\}, \{i_3, \xcancel{i_4}\}, \{\xcancel{i_5},\xcancel{i_6}\}, \{i_7\} \big\}$ \\ $\big\{ \{ \textcolor{mycolor}{i_3, i_2} \}, \{ \textcolor{mycolor}{i_7, i_1} \}, \{ i_6\textcolor{mycolor}{} \}, \{ i_4\textcolor{mycolor}{} \}, \{ i_5\textcolor{mycolor}{} \} \big\}$}& 
    \scalebox{1.3}{$\left( \begin{smallmatrix} m_1 & m_2 & m_3 & m_4 & m_5\\ i_1 & i_2 & i_3,i_4  & i_5,i_6 & i_7 \end{smallmatrix}  \right)$} & $\emptyset$& $ \emptyset $ \\\hline
\end{tabular}
\normalsize
\end{table}
\vspace{0.2cm}

At Step $1$, the EM mechanism starts with the permissible outcome $(E^1, A^1)=(I, \emptyset)$. 
Given $(E^1, A^1)$, major $m_5 \in M^{\mathrm{in+}}(E^1, A^1)$ is transfer-in expandable because $\bar{A}_{m_5} \neq \emptyset$ and $\mu_{m_5}(E^1, A^1) < \overline{q}_{m_5}$. 
Then, we go to step $1.1$. Major $m_5$ gives transfer-in eligibility to the student $\max_{\Pin_{m_5}} \bar{A}^1_{m_5}=i_5$. 
Set $A^2=\{i_5\}$, $E^2=E^1=I$, and go to Step $2$. 
We have a similar situation at Step $2$ and Step $3$ so that we omit the analysis. 

At Step $4$, given $(E^4, A^4)$ as shown in the table, there is no transfer-in expandable major, i.e., $M^{\mathrm{in+}}(E^4, A^4)=\emptyset$ and hence go to step $4.2$. 
Major $m_4 \in M^{\mathrm{out-}}(E^4, A^4)$ because $\mu_{m_4}(E^4, A^4) < \underline{q}_{m_4}$. 
Then $m_4$ revokes the transfer-out eligibility of agent $\min_{\Pout_{m_4}} E^4_{m_4}=i_6$. 
Set $E^5=E^4 \setminus \{i_6\}=I\setminus\{i_6\}$, $A^5=A^4$ and go to step $5$. 
We have a similar situation at Step $5$ and Step $6$ so that we omit the analysis. 

At Step $7$, given $(E^7, A^7)$ as shown in the table, we have $M^{\mathrm{in+}}(E^7, A^7)=\emptyset$ and $M^{\mathrm{out-}}(E^7, A^7)=\emptyset$ so that the EM mechanism stops. 

In summary, the outcome produced by the EM mechanism is $(E^7, A^7)$, where the transfer-out eligibility set is $E^7=\{i_1, i_2, i_3, i_7\}$ and the transfer-in eligibility set is $A^7=\{i_4, i_5, i_6\}$. 
That is, no student can transfer to their applied major because $E^7 \cap A^7 = \emptyset$.

\textit{According to Proposition \ref{prop_incom}, no major transition mechanism can be efficient and have eligibility maximization.} 
In this example, the outcome $(E^7, A^7)$ produced by the EM mechanism is a permissible EM outcome, but not Pareto efficient. It can be Pareto dominated by efficient outcomes $(E^*, A^*)$ and $(E^\diamond, A^\diamond)$, as stated previously. 

We now offer an insight into why a permissible EM outcome may not be Pareto efficient. 
We continue using Example \ref{Ex_incom} as a case in point. 
Since $(\underline{q}_{m_5}, \overline{q}_{m_5}) =(0,2)$ and we have only one transfer-out applicant and one transfer-in applicant, major $m_5 \in M^F(E, A)$ at any permissible EM outcome $(E, A)$. 
This implies that student $i_7$ has transfer-out eligibility and student $i_5$ has transfer-in eligibility at any permissible EM outcome $(E, A)$, including the result $(E^7, A^7)$ produced by the EM mechanism.

In Example \ref{Ex_incom}, in any permissible EM outcome where agent $i_5$ obtains transfer-in eligibility but does not receive transfer-out eligibility, agent $i_5$ acts as an interrupter and causes inefficiency in the permissible EM outcome. 
Specifically, agent $i_5$ prevents a swap between $i_4$ and $i_6$ in the following way. If they were to swap, then students $i_3$ and $i_4$ would obtain transfer-out eligibility from major $m_3$; students $i_5$ and $i_6$ would obtain transfer-out eligibility from major $m_4$; student $i_6$ would obtain transfer-in eligibility from major $m_3$; and student $i_4$ would obtain transfer-in eligibility from major $m_4$. 
As a result, $i_5$ would acquire both transfer-in and transfer-out eligibility, allowing $i_5$ to be assigned to major $m_5$. 
However, since $i_4$ and $i_6$ swapped their majors, only student $i_4$ would be assigned to $m_4$, which would violate $m_4$'s floor constraint. 
This situation prevents $i_6$ from obtaining transfer-out eligibility, thereby also violating $m_3$'s floor constraint and causing $i_4$ to fail to obtain transfer-out eligibility. 
Consequently, $i_5$ also does not receive transfer-out eligibility. 
In this case, agent $i_5$'s transfer-in eligibility is redundant. 
A Pareto improvement would involve removing $i_5$'s transfer-in eligibility, which would permit $i_4$ and $i_6$ to swap majors permissibly.
It is worth mentioning that an additional precondition for $i_4$ and $i_6$ swapping their majors permissibly is that student $i_3$ obtains transfer-out eligibility and does not have transfer-in eligibility. Otherwise, using the same logic, $i_3$ is an interrupter who also prevents the swap. 

Similarly, in any permissible EM outcome where agent $i_7$ obtains transfer-out eligibility but does not receive transfer-in eligibility, agent $i_7$ acts as an interrupter similarly to $i_5$. 
Specifically, agent $i_7$ prevents a swap between $i_1$ and $i_2$. 
A Pareto improvement would involve removing $i_7$'s transfer-out eligibility, which could allow $i_1$ and $i_2$ to swap majors permissibly. 
Notably, in this case, an additional precondition for $i_1$ and $i_2$ swapping their majors permissibly is that student $i_3$ obtains transfer-in eligibility and does not have transfer-out eligibility. Otherwise, using the same logic, $i_3$ is an interrupter who also prevents the swap.

More importantly, $i_3$ could act as an interrupter for both the swap between $i_4$ and $i_5$ and the swap between $i_1$ and $i_2$, depending on whether $i_3$ has transfer-out eligibility or transfer-in eligibility.\footnote{\ In this example, under any permissible outcome, student $i_3$ can obtain either transfer-in eligibility or transfer-out eligibility, but not both.} 
Specifically, 
when $i_3$ has only transfer-out eligibility, $i_3$ interrupts the swap between $i_1$ and $i_2$; and when $i_3$ has only transfer-in eligibility, $i_3$ interrupts the swap between $i_4$ and $i_6$.

In summary, this example highlights an important principle for achieving a Pareto improvement based on a permissible EM outcome: identifying interrupters and removing redundant transfer-in eligibility and transfer-out eligibility in a proper way.

\section{Efficiency-adjusted EM Mechanisms}

In this section, we introduce two efficiency-adjusted EM mechanisms in which the produced outcome can (weakly) Pareto dominate the input: a permissible EM outcome.  
The structure of this section is as follows. 
In the first part, we introduce the definitions of two types of exchangeable cycles: transfer-in exchangeable cycles and transfer-out exchangeable cycles, which can achieve a Pareto improvement in a permissible way. 
In the second part, we introduce a process for identifying transfer-in exchangeable cycles and discuss its properties.
Similarly, in the third part, we introduce a process for identifying transfer-out exchangeable cycles, and discuss its properties.
In the last part, we introduce the efficiency-adjusted EM mechanisms based on the previously defined processes. 

\subsection{A Transfer-in Exchangeable Cycle and A Transfer-out Exchangeable Cycle}

In a major transition program, a student $i$ can transfer her major successfully only when the student $i$ obtains both the transfer-out eligibility of $\omega_i$ and the transfer-in eligibility of $\alpha_i$. 
In contrast, if student $i$ obtains only transfer-out eligibility or transfer-in eligibility, then $i$ fails to transfer his/her major. 

We illustrate that students who obtain only one-side eligibility may improve their welfare permissibly through two types of exchangeable cycle: transfer-in exchangeable cycles and transfer-out exchangeable cycles. 
Such methods of improvement are an essential component of the efficient mechanisms proposed in this paper.

\subsubsection{A Transfer-in Exchangeable Cycle}

Fix a permissible outcome $(E, A)$. 
We show that students who obtain only the transfer-out eligibility can search to possibly improve their welfare by exchanging their majors in a permissible way. 
We say that a major $m$ is \textbf{transfer-in exchangeable} if there exists an applicant $i$ who obtains transfer-out eligibility from $\omega_i$ but does not have transfer-in eligibility from $m$, i.e.,
$\bar{A}_m \cap E \neq \emptyset$. 
Notably, students who obtain only the transfer-out eligibility imply that their applied majors are transfer-in exchangeable. 
Consider the following pointing process: 

\begin{itemize}
    \item \textit{Each transfer-in exchangeable major points to the highest priority transfer-in applicant among its transfer-in applicants who have only transfer-out eligibility; each applicant who is pointed by a transfer-in exchangeable major points to their initial major.}
\end{itemize}

If there exists a cycle, we call it \textbf{a transfer-in exchangeable cycle at $(E, A)$}. We write $\Cin=(m_0,i_1, m_1, \ldots,i_L)$ to represent a transfer-in exchangeable cycle, where $m_l=\omega_{i_l}$ and $m_{l-1}=\alpha_{i_{l}}$ for each $1 \leq l\leq L$ and  $m_{L}=m_0$.\footnote{\ The pointing process of the transfer-in exchangeable cycle $\Cin$ is that $\alpha_{i_1}=\omega_{i_L}=m_0 \to i_1 \to \omega_{i_1}=\alpha_{i_2}=m_1 \to ... \to i_m \to \omega_{i_m}=\alpha_{i_1}=m_0$. Slightly abusing notation, we use $\Cin$ to represent the set of students and majors in this cycle.}

Each transfer-in exchangeable cycle can be implemented via a new permissible outcome, such that each student in the cycle can successfully transfer their major. 
This is achieved by additionally giving transfer-in eligibility to every transfer-in applicant who has a weakly higher transfer-in priority than the pointed student in the cycle. 
Moreover, the new outcome Pareto dominates the initial permissible outcome. 
Formally,

\begin{proposition}\label{Prop_TiEc}
    Given a permissible outcome $(E, A)$, any transfer-in exchangeable cycle $\Cin=(m_0,i_1,\ldots,i_L)$ at $(E, A)$ is implementable via the permissible outcome $(E', A')$, where $E'=E$ and $A'=A \cup (\bigcup_{m_l \in \Cin} \{i \in \bar{A}_{m_l} \vert i \Pineq_{m_l} i_{l+1}\})$ ($l \in \{0,...,L-1\}$). Moreover, the outcome $(E', A')$ Pareto dominates the outcome $(E, A)$.  
\end{proposition} 

We use an example to illustrate the Pareto improvement via a transfer-in exchangeable cycle.\footnote{\ Without loss of generality, we start with a permissible EM outcome. The analysis also holds for any permissible outcome.}

\begin{example}\label{Ex_transcycle}
Consider two students $i_1, i_2$ and two majors $m_1, m_2$ such that $\omega_{i_1}=\alpha_{i_2}=m_1$ and $\omega_{i_2}=\alpha_{i_1}=m_2$. The distributional constraints of $m_1$ and $m_2$ are $(\underline{q}_{m_1}, \overline{q}_{m_1})=(1, 1)$ and $(\underline{q}_{m_2}, \overline{q}_{m_2})=(1, 1)$, respectively. 
The transfer-out priority of major $m_1$ is $\Pout_{m_1}: i_1$ and that of major $m_2$ is $\Pout_{m_2}: i_2$. 
The transfer-in priority of major $m_1$ is $\Pin_{m_1}: i_2$ and that of major $m_2$ is $\Pin_{m_2}: i_1$. 
There are three permissible EM outcomes, which are $(E, A)=(\{i_1, i_2\}, \emptyset)$, $(E', A')=(\emptyset, \{i_1, i_2\})$, and $(E^*, A^*)=(\{i_1, i_2\}, \{i_1, i_2\})$. 
The outcome $(E^*, A^*)$ is Pareto efficient. 
\end{example}

\begin{figure}[!ht]
    \begin{center}
    \begin{subfigure}[t]{0.45\textwidth}
    \centering
    \begin{tikzpicture}[>=stealth, line width=0.8pt,scale=0.8]
    \draw[rounded corners] (-4,1) -- (-4,-1) -- (-2,-1) -- (-2,1) -- cycle;
    \draw[rounded corners] (-1,1) -- (-1,-1) -- (1,-1) -- (1,1) -- cycle;

    \draw[white, thick] (-3.7,-1) -- (-3.3,-1); 
    \draw[white, thick] (-2.3,-1) -- (-2.7,-1); 
    \draw[white, thick] (-0.3,-1) -- (-0.7,-1); 
    \draw[white, thick] (0.3,-1) -- (0.7,-1); 

    \node at (-3.5,-1) {$E$}; 
    \node at (-2.5,-1) {$A$}; 
    \node at (-0.5,-1) {$E$}; 
    \node at (0.5,-1) {$A$}; 
    \node at (-3, 1.5) {$m_1$};
    \node at (0, 1.5) {$m_2$};
    \node at (-3.5,0) {$i_1$};
    \node at (-2.5,0) {$\textcolor{mycolor}{i_2}$};
    \node at (-0.5,0) {$i_2$};
    \node at (0.5,0) {$\textcolor{mycolor}{i_1}$};

    \draw[->,red] (-3.5,0.15) -- (-3.1,1.4) ; 
    \draw[->,red] (-2.9,1.4) -- (-2.5,0.15) ; 
    \draw[->,red] (-0.5,0.15) -- (-0.1,1.4) ; 
    \draw[->,red] (0.1,1.4) -- (0.5,0.15) ; 

    \draw[dashed] (-3,1) to (-3,-1);
    \draw[dashed] (0,1) to (0,-1);
    \draw[dashed] (-2.4,0) to (-0.6,0);
    \draw[dashed] (-3.5,-0.1) to[bend right=30] (0.5,-0.1);
    \end{tikzpicture}
    \subcaption{\footnotesize{A transfer-in exchangeable cycle}}\label{Fig_cyclesa} 
    \end{subfigure} 
    \begin{subfigure}[t]{0.45\textwidth}
    \centering
    \begin{tikzpicture}[>=stealth, line width=0.8pt,scale=0.8]
    \draw[rounded corners] (-4,1) -- (-4,-1) -- (-2,-1) -- (-2,1) -- cycle;
    \draw[rounded corners] (-1,1) -- (-1,-1) -- (1,-1) -- (1,1) -- cycle;

    \draw[white, thick] (-3.7,-1) -- (-3.3,-1); 
    \draw[white, thick] (-2.3,-1) -- (-2.7,-1); 
    \draw[white, thick] (-0.3,-1) -- (-0.7,-1); 
    \draw[white, thick] (0.3,-1) -- (0.7,-1); 

    \node at (-3.5,-1) {$E$}; 
    \node at (-2.5,-1) {$A$}; 
    \node at (-0.5,-1) {$E$}; 
    \node at (0.5,-1) {$A$}; 
    \node at (-3, 1.5) {$m_1$};
    \node at (0, 1.5) {$m_2$};
    \node at (-3.5,0) {$\xcancel{i_1}$};
    \node at (-2.5,0) {$i_2$}; 
    \node at (-0.5,0) {$\xcancel{i_2}$};
    \node at (0.5,0) {$i_1$};

    \draw[->,blue]  (-3.1,1.4) -- (-3.5,0.15)  ; 
    \draw[->,blue] (-2.5,0.15) -- (-2.9,1.4) ; 
    \draw[->,blue] (-0.1,1.4) -- (-0.5,0.15)  ; 
    \draw[->,blue] (0.5,0.15) -- (0.1,1.4) ; 

    \draw[dashed] (-3,1) to (-3,-1);
    \draw[dashed] (0,1) to (0,-1);
    \draw[dashed] (-2.4,0) to (-0.6,0);
    \draw[dashed] (-3.5,-0.1) to[bend right=30] (0.5,-0.1);
    \end{tikzpicture}
    \caption{\footnotesize{A transfer-out exchangeable cycle}}\label{Fig_cyclesb} 
    \end{subfigure}
    \caption{Two types of exchangeable cycles }\label{Fig_cycles}
    \end{center}
    \vspace{-1.5em}
\end{figure}
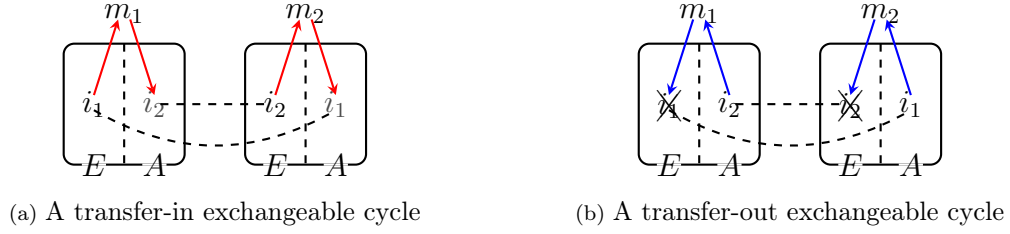

Consider the permissible EM outcome 
$(E, A)=(\{i_1, i_2\}, \emptyset)$. 
It is not Pareto efficient because it can be Pareto dominated by the outcome $(E^*, A^*)=(\{i_1, i_2\}, \{i_1, i_2\})$. 
Given $(E, A)=(\{i_1, i_2\}, \emptyset)$, both major $m_1$ and major $m_2$ are transfer-in exchangeable. 
Figure~\ref{Fig_cycles}~(a) shows the pointing process for finding a transfer-in exchangeable cycle.\footnote{\ In this paper, we use a box to represent the information of transfer-out and transfer-in applicants in a major. The left side of each box displays the major's transfer-out applicants. If there is more than one applicant, they are ordered by its transfer-out priority, with the highest priority applicant at the top, followed by the second highest, and so on. Similarly, the left side of each box displays the major's transfer-in applicants.}
In particular, major $m_1$ points to the transfer-in applicant $i_2$ with the highest transfer-in priority, who has only transfer-out eligibility; student $i_2$ points to the current major $m_2$; major $m_2$ points to the transfer-in applicant $i_1$ with the highest transfer-in priority, who has only transfer-out eligibility; student $i_1$ points to the current major $m_1$. 
Therefore, we have the transfer-in exchangeable cycle $\Cin=(m_1, i_2, m_2, i_1)$. 
In this case, $i_1$ and $i_2$ can swap their majors without violating the priorities and distributional constraints of all majors, via additionally distributing transfer-in eligibility to students $i_1$ and $i_2$, resulting in the unique Pareto efficient outcome $(E^*, A^*)$.

\subsubsection{A Transfer-out Exchangeable Cycle}
Fix a permissible outcome $(E, A)$. 
Similar to the analysis in the previous subsection, we show that students who obtain only transfer-in eligibility can search for ways to potentially improve their welfare by exchanging their majors without violating permissibility. 
We say that a major $m$ is \textbf{transfer-out exchangeable} if there exists an applicant $i$ who obtains transfer-in eligibility from $\alpha_m$ but does not have transfer-out eligibility from $m$, i.e., $(\omega_m \cap A) \setminus E_m \neq \emptyset$. 
Notably, if students obtain only transfer-in eligibility, this implies that their initial majors are transfer-out exchangeable. 
Consider the following pointing process: 

\begin{itemize}
    \item \textit{Each transfer-out exchangeable major points to the highest priority transfer-out applicant among its transfer-out applicants who have only transfer-in eligibility; each applicant who is pointed by a transfer-out exchangeable major points to their applied major.} 
\end{itemize}

If there exists a cycle, we call it \textbf{a transfer-out exchangeable cycle at $(E, A)$}. We write $\Cout=(m_1,i_1,m_2,...,i_L)$ to represent a transfer-out exchangeable cycle, where $m_l=\omega_{i_l}$ and $m_{l+1}=\alpha_{i_l}$ for each $1 \leq l \leq L$ and $m_{L+1}=m_1$.\footnote{\ The pointing process of the transfer-in exchangeable cycle $\Cout$ is that $\omega_{i_1}=m_1 \to i_1 \to \alpha_{i_1}=\omega_{i_2} \to ...\to i_m \to \alpha_{i_m}=\omega_{i_1}=m_1$. Slightly abusing notation, we use $\Cout$ to represent the set of students and majors in this cycle.}

Each transfer-out exchangeable cycle is implemented via a new permissible outcome, such that every transfer-out applicant in the cycle transfers their major successfully. 
This is achieved by additionally giving transfer-out eligibility to every student who has a weakly higher transfer-out priority than the pointed student in the cycle. Moreover, the new outcome Pareto dominates the initial permissible outcome. 
Formally, 

\begin{proposition}\label{Prop_ToEc}
    Given a permissible outcome $(E, A)$, any transfer-out exchangeable cycle $\Cout=(m_1,i_1,...,i_L)$ is implementable via the permissible outcome $(E', A')$, where $E'=E \cup (\bigcup_{m_l \in \Cout} \{i \in \bar{E}_{m_l} \vert i \Pouteq_{m_l} i_{l}\})$ ($l \in \{1,...,L\}$)and $A'=A$. 
    Moreover, the outcome $(E', A')$ Pareto dominates the outcome $(E, A)$. 
\end{proposition}

We continue using Example \ref{Ex_transcycle} to illustrate the Pareto improvement via a transfer-out exchangeable cycle. 
Consider the permissible EM outcome $(E', A)=(\emptyset, \{i_1, i_2\})$.
It is not Pareto efficient because it can be Pareto dominated by the outcome $(E^*, A^*)=(\{i_1, i_2\}, \{i_1, i_2\})$. 
As shown in Figure~\ref{Fig_cycles}~(b), we have a transfer-out exchange cycle $\Cout=(m_1, i_1, m_2, i_2)$. 
In this case, $i_1$ and $i_2$ can swap their majors without violating the priorities and distributional constraints of all majors, via additionally distributing transfer-out eligibility to students $i_1$ and $i_2$, resulting in the unique Pareto efficient outcome $(E^*, A^*)$.

Given a permissible EM outcome $(E, A)$, by Lemma \ref{Le_AT}, each student $i \in I$ has either $i \in E$ or $i \in A$, or both. 
By Proposition \ref{Prop_AT2}, we have the following Corollary.

\begin{corollary}\label{Co_EM}
Given a permissible EM outcome $(E, A)$, any permissible outcome $(E', A')$ Pareto dominates $(E, A)$ through either transfer-in exchangeable cycles, transfer-out exchangeable cycles, or both.
\end{corollary}

\subsection{A Process for Finding Transfer-in Exchangeable Cycles} 

In this subsection, we propose a process for finding transfer-in exchangeable cycles, referred to as \textit{the TiE process}.

The TiE process starts with a permissible outcome. 
To identify all transfer-in exchangeable cycles that can be implemented permissibly, the process classifies the remaining majors at each step into two types, via a pointing process. 
The first type consists of majors that can no longer participate in any transfer-in exchangeable cycle. These majors will revoke the transfer-out eligibility of agents with lower priority than the lowest-priority agent who obtains both transfer-out and transfer-in eligibility, and will be removed once identified.  
The second type includes majors that can possibly engage in a transfer-in exchangeable cycle. 
If at any step there is no first type major, this indicates  the existence of at least one transfer-in exchangeable cycle. 
In this situation, all students involved in the cycle can swap their majors among themselves without violating each major's dual priority and distributional constraints. 
The process stops once all majors are removed. 
A formal description of the TiE process is as follows. 

\begin{center}
\noindent\textbf{A process for finding transfer-in exchangeable cycles $f^{TiE}$} 
\end{center}

\begin{itemize} 
    \item[Step $0$] Given a permissible outcome $(E, A)$, let $E^1 = E$, $A^1 = A$ and $J^1 = M$.  
        
    \item[Step $k$] ($k \geq 1$) If $J^k = \emptyset$, the process stops. Otherwise, do the following pointing process: 
    
    \begin{itemize}
        \item Each major $m \in J^k$ points to the highest priority transfer-in applicant $i^{* \leftarrow m} = \max_{\Pin_m} (\bar{A}_m^k \cap E^k)$ among the transfer-in applicants who have only transfer-out eligibility (if any). 
        Each pointed applicant points to their current major. 
    \end{itemize}

    Let $N^k$ be the set of majors that either have no applicant to point, or the pointed applicant $i^{* \leftarrow m}$'s current major does not belong to the set $J^k$, i.e., $\omega_{i^{* \leftarrow m}} \notin J^k$.
   
    \begin{itemize}
        \item[(a)] If $N^k \neq \emptyset$, each major $m \in N^k$ evokes the transfer-out eligibility of every applicant from set $T^{\mathrm{out}}_m = \{i \in E^k_m \mid i^* \Pout_m i \}$ (if any), where $i^* = \min_{\Pout_m} (E^k_m \cap A^k)$. 
        Set $E^{k+1} = E^k \setminus (T^{\mathrm{out}}_m)_{m \in N^k}$, $A^{k+1} = A^k$ and $J^{k+1} = J^{k} \setminus N^k$. 
        Let $k=k+1$ and go to step $k$.

        \item[(b)] If $N^k = \emptyset$, there exists at least one transfer-in exchangeable cycle. 
        Let $C^k$ be the set of cycles. 
        Each major $m \in C^k$ distributes transfer-in eligibility to each applicant from the set $T^{\mathrm{in}}_m = \{i \in \bar{A}^k_m \mid i \Pineq_m i^{* \leftarrow m} \}$. 
        Set $A^{k+1} = A^{k} \cup (T^{\mathrm{in}}_m)_{m \in C^k}$, $E^{k+1} = E^{k}$, $J^{k+1}=J^k$. 
        Let $k=k+1$ and go to step $k$. 

    \end{itemize}
\end{itemize}

Due to the finite set of applicants and majors, it is clear that the TiE process terminates in finite steps. 
Given a permissible outcome $(E, A)$, let $f^{TiE}(E, A)=(E^{TiE}, A^{TiE})$ be the outcome produced by the TiE process.

\begin{remark}\label{Rm_ParetoImprove}
Let $K$ be the last step of the TiE process. According to the TiE process, the following remarks hold true: 

(a) Every major $m$ involved in a cycle is transfer-in exchangeable;

(b) $E^{k+1} \subseteq E^{k}$ and $A^{k} \subseteq A^{k+1}$ hold true for each step $k < K$;

(c) $\min_{\Pout_m} E^{TiE} \in A^{TiE}$ and $\min_{\Pin_m} A^{TiE} \in E^{TiE}$ hold true for each major $m \in M$.
\end{remark}

Remark \ref{Rm_ParetoImprove}(a) indicates that every cycle formed in the TiE process is a transfer-in exchangeable cycle. 
Remark \ref{Rm_ParetoImprove}(b) implies that during the TiE process, the transfer-out eligibility set is (weakly) monotonically decreasing and the transfer-in eligibility set is (weakly) monotonically increasing. 
Remark \ref{Rm_ParetoImprove}(c) shows that for each major, the lowest priority agent who obtains transfer-out eligibility (transfer-in eligibility) also obtains transfer-in eligibility (transfer-out eligibility) at outcome $f^{TiE}(E, A)$. 

According to the TiE process and proposition \ref{Prop_TiEc}, Lemma~\ref{Le_TiEpermissible} holds true. 

\begin{lemma}\label{Le_TiEpermissible}
Given a permissible outcome $(E, A)$, the outcome $f^{TiE}(E, A)$ is permissible. 
\end{lemma}

\begin{lemma}\label{Le_TiE}
Given a permissible outcome $(E, A)$, the outcome $f^{TiE}(E, A)$ produced by the TiE process has either $\mu {(f^{TiE}(E, A))}=\mu {(E, A)}$ or Pareto dominates the outcome $(E, A)$. 
\end{lemma}

Next, we highlight the important role of the TiE process in designing an efficient mechanism. 
Given a permissible EM outcome $(E, A)$, we say that student $i$ is \textbf{not Pareto improvable at $(E, A)$} if for every permissible outcome $(E', A')$ that Pareto dominates the outcome $(E, A)$, $\mu_i (E', A')= \mu_i (E, A)$. 

\begin{proposition}\label{Prop_TiEmain}
Given a permissible EM outcome $(E, A)$, let $f^{TiE}(E, A)=(E^{TiE}, A^{TiE})$ be the outcome produced by the TiE process. 
Then, every student $i \in I^O (E,A)$ is not Pareto improvable at $(E^{TiE}, A^{TiE})$.  
\end{proposition}

Proposition \ref{Prop_TiEmain} indicates that every student assigned to overdemanded majors at $(E,A)$ cannot achieve a Pareto improvement in any permissible outcome that Pareto dominates the TiE outcome $(E^{TiE}, A^{TiE})$. 
It is worth mentioning that the proposition only holds when the input of the TiE process is a permissible EM outcome.

We continue using Example \ref{Ex_incom} to illustrate how the TiE process works.

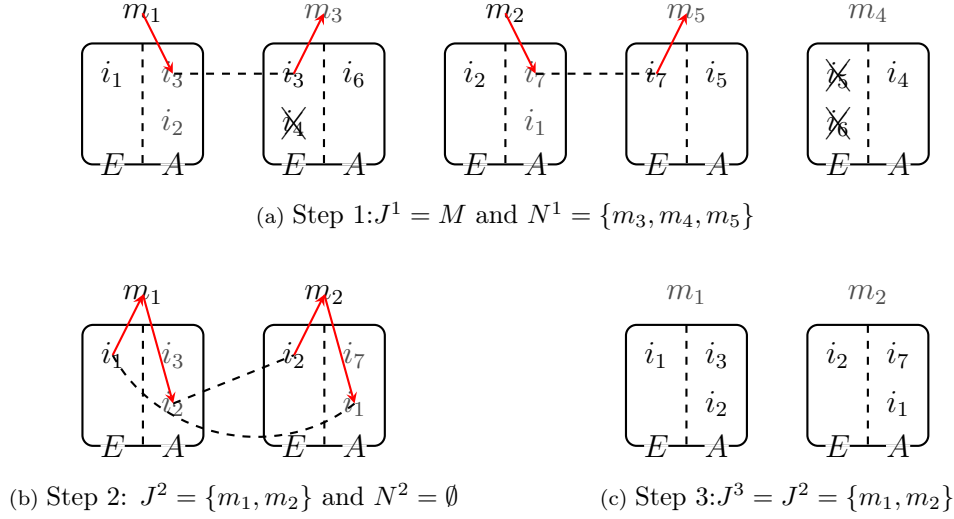
\begin{figure}[htbp]
    \begin{subfigure}[t]{\textwidth}
    \centering
    \begin{tikzpicture}[>=stealth, line width=0.8pt,scale=0.8]
        \foreach \i in {0, 1, 2, 3, 4} {
            \draw[rounded corners] ({-4 + \i * 3}, 1) -- ({-4 + \i * 3}, -1) -- ({-2 + \i * 3}, -1) -- ({-2 + \i * 3}, 1) -- cycle;
        }

        \draw[white, thick] (-3.7,-1) -- (-3.3,-1); 
        \draw[white, thick] (-2.7,-1) -- (-2.3,-1); 
        \draw[white, thick] (-0.7,-1) -- (-0.3,-1); 
        \draw[white, thick] (0.3,-1) -- (0.7,-1); 
        \draw[white, thick] (2.3,-1) -- (2.7,-1); 
        \draw[white, thick] (3.3,-1) -- (3.7,-1); 
        \draw[white, thick] (5.3,-1) -- (5.7,-1); 
        \draw[white, thick] (6.3,-1) -- (6.7,-1); 
        \draw[white, thick] (8.3,-1) -- (8.7,-1); 
        \draw[white, thick] (9.3,-1) -- (9.7,-1); 

        \node at (-3.5, -1) {$E$};
        \node at (-2.5,-1) {$A$};
        \node at (-0.5, -1) {$E$};
        \node at (0.5,-1) {$A$};
        \node at (2.5, -1) {$E$};
        \node at (3.5,-1) {$A$};
        \node at (5.5, -1) {$E$};
        \node at (6.5,-1) {$A$};
        \node at (8.5, -1) {$E$};
        \node at (9.5,-1) {$A$};

        \node at (-3,1.5) {$m_1$};
        \node at (0,1.5) {$\textcolor{mycolor}{m_3}$};
        \node at (3,1.5) {$m_2$};
        \node at (9,1.5) {$\textcolor{mycolor}{m_4}$};
        \node at (6,1.5) {$\textcolor{mycolor}{m_5}$};

        \node at (-3.5, 0.5) {$i_1$};
        \node at (-2.5, 0.5) {$\textcolor{mycolor}{i_3}$};
        \node at (-2.5, -0.3) {$\textcolor{mycolor}{i_2}$};
        \node at (-0.5, 0.5) {$i_3$};
        \node at (-0.5, -0.3) {$\xcancel{i_4}$};
        \node at (0.5, 0.5) {$i_6$};
        \node at (2.5, 0.5) {$i_2$};
        \node at (3.5, 0.5) {$\textcolor{mycolor}{i_7}$};
        \node at (3.5,-0.3) {$\textcolor{mycolor}{i_1}$};
        \node at (8.5, 0.5) {$\xcancel{i_5}$};
        \node at (8.5, -0.3) {$\xcancel{i_6}$};
        \node at (9.5, 0.5) {$i_4$};
        \node at (5.5, 0.5) {$i_7$};
        \node at (6.5, 0.5) {$i_5$};
    
    \draw[dashed] (-3,1) to (-3,-1);
    \draw[dashed] (0,1) to (0,-1);
    \draw[dashed] (3,1) to (3,-1);
    \draw[dashed] (6,1) to (6,-1);
    \draw[dashed] (9,1) to (9,-1);
    \draw[dashed] (-0.5,0.5) to (-2.5,0.5);
    \draw[dashed] (3.5, 0.5) to (5.5, 0.5);

    \draw[->, red] (-3,1.5) -- (-2.5,0.5);
    \draw[->, red] (3,1.5) -- (3.5, 0.5);
    \draw[->, red] (-0.5,0.5) -- (0,1.5);
    \draw[->, red] (5.5, 0.5) -- (6,1.5);    
\end{tikzpicture}
\caption{\footnotesize{Step 1:$J^1=M \text{  and } N^1=\{m_3,m_4,m_5\}$}}
\end{subfigure}

\vspace{0.5em}

\begin{center}

\begin{subfigure}[t]{0.4\textwidth}
\centering
\begin{tikzpicture}[>=stealth, line width=0.8pt,scale=0.8]
    \foreach \i in {0, 1} {
        \draw[rounded corners] ({-4 + \i * 3}, 1) -- ({-4 + \i * 3}, -1) -- ({-2 + \i * 3}, -1) -- ({-2 + \i * 3}, 1) -- cycle;
    }

    \draw[white, thick] (-3.7,-1) -- (-3.3,-1); 
    \draw[white, thick] (-2.7,-1) -- (-2.3,-1); 
    \draw[white, thick] (-0.7,-1) -- (-0.3,-1); 
    \draw[white, thick] (0.3,-1) -- (0.7,-1); 

    \node at (-3.5, -1) {$E$};
    \node at (-2.5,-1) {$A$};
    \node at (-0.5, -1) {$E$};
    \node at (0.5,-1) {$A$};

    \node at (-3,1.5) {$m_1$};
    \node at (0,1.5) {$m_2$};
    
    \node at (-3.5, 0.5) {$i_1$};
    \node at (-2.5, 0.5) {$\textcolor{mycolor}{i_3}$};
    \node at (-2.5, -0.3) {$\textcolor{mycolor}{i_2}$};
    \node at (-0.5, 0.5) {$i_2$};
    \node at (0.5, 0.5) {$\textcolor{mycolor}{i_7}$};
    \node at (0.5, -0.3) {$\textcolor{mycolor}{i_1}$};
    
    \draw[dashed] (-3,1) to (-3,-1);
    \draw[dashed] (0,1) to (0,-1);
    \draw[dashed] (-3.5,0.5) to[bend right=50] (0.5,-0.3);
    \draw[dashed] (-2.5,-0.3) to (-0.5, 0.5);

    \draw[->, red] (-3,1.5) -- (-2.5,-0.3);
    \draw[->, red] (0,1.5) -- (0.5,-0.3);
    \draw[->, red] (-3.5,0.5) -- (-3,1.5);
    \draw[->, red] (-0.5, 0.5) -- (0,1.5);    
\end{tikzpicture}
\caption{\footnotesize{Step 2: $J^2=\{m_1,m_2\} \text{  and } N^2=\emptyset$}}
\end{subfigure}
\quad
\begin{subfigure}[t]{0.4\textwidth}
\centering
\begin{tikzpicture}[>=stealth, line width=0.8pt, scale=0.8]
    \foreach \i in {0, 1} {
        \draw[rounded corners] ({-4 + \i * 3}, 1) -- ({-4 + \i * 3}, -1) -- ({-2 + \i * 3}, -1) -- ({-2 + \i * 3}, 1) -- cycle;
    }

    \draw[white, thick] (-3.7,-1) -- (-3.3,-1); 
    \draw[white, thick] (-2.7,-1) -- (-2.3,-1); 
    \draw[white, thick] (-0.7,-1) -- (-0.3,-1); 
    \draw[white, thick] (0.3,-1) -- (0.7,-1); 
    
    \node at (-3.5, -1) {$E$};
    \node at (-2.5,-1) {$A$};
    \node at (-0.5, -1) {$E$};
    \node at (0.5,-1) {$A$};

    \node at (-3,1.5) {$\textcolor{mycolor}{m_1}$};
    \node at (0,1.5) {$\textcolor{mycolor}{m_2}$};

    \node at (-3.5, 0.5) {$i_1$};
    \node at (-2.5, 0.5) {$i_3$};
    \node at (-2.5, -0.3) {$i_2$};
    \node at (-0.5, 0.5) {$i_2$};
    \node at (0.5, 0.5) {$i_7$};
    \node at (0.5, -0.3) {$i_1$};
    
    \draw[dashed] (-3,1) to (-3,-1);
    \draw[dashed] (0,1) to (0,-1);

\end{tikzpicture}
\caption{\footnotesize{Step 3:$J^3=J^2=\{m_1,m_2\}$}}
\end{subfigure}
\caption{The TiE process in Example \ref{Ex_incom} given the permissible EM outcome $(E^7, A^7)$ as input}\label{Fig_Ex_TiE}
\end{center}
\vspace{-1.5em}
\end{figure}

Take the permissible EM outcome produced by the EM mechanism as an input so that 
set $(E^1, A^1)=(\{i_1,i_2,i_3,i_7\}, \{i_4,i_5,i_6\})$ and $J^1=M$. 

At Step $1$, major $m_1$ points to $i_3$, and major $m_2$ points to $i_7$; $i_3$ points to the initial major $m_3$ and $i_7$ points to the initial major $m_5$. 
As shown in Figure \ref{Fig_Ex_TiE}(a), we have $N^1=\{m_3,m_4,m_5\} \neq \emptyset$ and hence go to process (a). 
In details, major $m_3$ evokes $i_3$'s transfer-out eligibility; major $m_5$ evokes $i_7$'s transfer-out eligibility. 
Let $(E^2, A^2)=(\{i_1,i_2\}, \{i_4,i_5,i_6\})$, $J^2=J^1\setminus N^1=\{m_1,m_2\}$, and go to Step 2.

At Step $2$, major $m_1$ points to $i_2$, and major $m_2$ points to $i_1$; 
$i_2$ points to the initial major $m_2$, and $i_1$ points to the initial major $m_1$. 
As shown in Figure \ref{Fig_Ex_TiE}(b), we have $N^2=\emptyset$ and hence go to process (b). 
There exists a transfer-in exchangeable cycle $C^{\mathrm{in}}=(m_1, i_2, m_2, i_1, m_1)$. 
Hence, major $m_1$ gives transfer-in eligibility to student $i_2$, and student $i_3$ who does not have transfer-in eligibility but has a higher transfer-in priority than $i_2$; 
major $m_2$ gives transfer-in eligibility to both student $i_1$ and student $i_7$ who does not have transfer-in eligibility but has higher transfer-in priority than $i_1$.
Let $(E^3, A^3)=(\{i_1,i_2\}, \{i_1, i_2, i_3,i_4,i_5,i_6, i_7\})$, $J^3=J^2=\{m_1,m_2\}$, and go to Step 3.

At Step 3, neither major $m_1$ nor major $m_2$ have a transfer-in applicant without transfer-in eligibility. 
As shown in Figure \ref{Fig_Ex_TiE}(c), $N^3=\{m_1,m_2\}$ and hence go to process (a). 
Since $T^{out}_{m_1}=T^{out}_{m_2}=\emptyset$, set $(E^4, A^4)=(\{i_1,i_2\}, \{i_1, i_2, i_3,i_4,i_5,i_6, i_7\})$, $J^4=\emptyset$, and go to Step 4.

At Step $4$, since $J^4=\emptyset$, the process stops. 

In summary, the outcome produced by the TiE process is $f^{TiE}(E, A)=(E^*, A^*)=(\{i_1,i_2\}, \{i_1, i_2, i_3,i_4,i_5,i_6, i_7\})$ and it is a permissible and Pareto efficient outcome.

\subsection{A Process for Finding Transfer-out Exchangeable Cycles}

In this subsection, we propose a process for finding transfer-out exchangeable cycles, referred to as \textit{the ToE process}.

The ToE process starts with a permissible outcome. 
To identify all transfer-out exchangeable cycles that can be implemented permissibly, the process classifies the remaining majors at each step into two types, via a pointing process. 
The first type consists of majors that can no longer participate in any transfer-out exchangeable cycle. 
These majors will revoke the transfer-in eligibility of agents with lower priority than the lowest-priority agent who obtains both transfer-out and transfer-in eligibility, and will be removed once identified. 
The second type includes majors that can possibly engage in a transfer-out exchangeable cycle. 
If at any step there is no first type major, this indicates  the existence of at least one transfer-out exchangeable cycle. 
In this situation, all students involved in the cycle can swap their majors among themselves without violating each major's dual priority and distributional constraints. 
The process stops once all majors are removed. 
A formal description of the ToE process is as follows.

\begin{center}
\noindent\textbf{A process for finding transfer-out exchange cycles $f^{ToE}$} 
\end{center}
\begin{itemize} 
    \item[Step $0$] Given a permissible outcome $(E, A)$, let $E^1=E$, $A^1=A$ and $J^1=M$. 
        
    \item[Step $k$] ($k \geq 1$) If $J^k = \emptyset$, the process stops. Otherwise, do the following pointing process: 

    \begin{itemize}
        \item Each major $m \in J^k$ points to the highest priority transfer-out applicant $i^{* \leftarrow m} = \max_{\Pout_m} (\bar{E}^k_m \cap A^k)$ among the transfer-out applicants who have only transfer-in eligibility (if any). 
        Each pointed applicant points to their applied major. 
    \end{itemize}
    
    Let $N^k$ be the set of majors that have no applicant to point, or the pointed applicant $i^{* \leftarrow m}$'s applied major does not belong to the set $J^k$, i.e., $\alpha_{i^{* \leftarrow m}} \notin J^k$. 

    \begin{itemize}
        \item[(a)] If $N^k \neq \emptyset$, each major $m \in N^k$ evokes the transfer-in eligibility of every applicant from set $T^{\mathrm{in}}_m=\{i \in A^k_m \vert i^* \Pin_m i \}$, where $i^* = \min_{\Pin_m} (A^k_m \cap E^k)$. 
        Set $A^{k+1}=A^k \setminus (T^{\mathrm{in}}_m)_{m \in N^k}$, $E^{k+1}=E^k$ and $J^{k+1}=J^{k} \setminus N^k$. 
        Let $k=k+1$ and go to step $k$.  

        \item[(b)] If $N^k = \emptyset$, there exists at least one transfer-out exchangeable cycle. 
        Let $C^k$ be the set of cycles. 
        Each major $m \in C^k$ in the cycle distributes transfer-out eligibility to each applicant from the set $T^{\mathrm{out}}_m=\{i \in \bar{E}_m^k \vert i \succeq^{\mathrm{out}}_m i^{* \leftarrow m}\}$. 
        Set $E^{k+1} = E^{k} \cup (T^{\mathrm{out}}_m)_{m \in C^k}$, $A^{k+1} = A^{k}$, $J^{k+1}=J^k$. 
        Let $k=k+1$ and go to step $k$. 
    \end{itemize}
\end{itemize}

Due to the finite set of applicants and majors, it is clear that the ToE process terminates in finite steps. 
Given a permissible outcome $(E, A)$, let $f^{ToE}(E, A)=(E^{ToE}, A^{ToE})$ be the outcome produced by the ToE process. 

\begin{remark}\label{Rm_ParetoImprove2}
Let $K$ be the last step of the ToE process. According to the ToE process, the following remarks hold true: 

(a) Every major $m$ involved in a cycle is transfer-out exchangeable; 

(b) $E^{k} \subseteq E^{k+1}$ and $A^{k+1} \subseteq A^{k}$ hold true for each step $k < K$; 

(c) $\min_{\Pout_m} E^{ToE} \in A^{ToE}$ and $\min_{\Pin_m} A^{ToE} \in E^{ToE}$ hold true for each major $m \in M$. 
\end{remark}

Remark \ref{Rm_ParetoImprove2}(a) indicates that 
every cycle formed in the ToE process is a transfer-out exchangeable cycle. 
Remark \ref{Rm_ParetoImprove2}(b) implies that during the ToE process, the transfer-out eligibility set is (weakly) monotonically increasing and the transfer-in eligibility set is (weakly) monotonically decreasing. 
Remark \ref{Rm_ParetoImprove2}(c) shows that for each major, the lowest priority agent who obtains transfer-out eligibility (transfer-in eligibility) also obtains transfer-in eligibility (transfer-out eligibility) at outcome $(E^{ToE}, A^{ToE})$.

According to the ToE process and proposition \ref{Prop_ToEc}, Lemma~\ref{Le_ToEpermissible} holds true. 

\begin{lemma}\label{Le_ToEpermissible}
Given a permissible outcome $(E, A)$, the outcome $f^{ToE}(E, A)$ is permissible. 
\end{lemma}

\begin{lemma}\label{Le_ToE}
Given a permissible outcome $(E, A)$, the outcome $f^{ToE}(E, A)$ produced by the ToE process has either $\mu {(f^{ToE}(E, A))}=\mu {(E, A)}$ or Pareto dominates the outcome $(E, A)$. 
\end{lemma}

Similarly, we show the important role of the ToE process in designing an efficient mechanism.

\begin{proposition}\label{Prop_ToEmain}
Given a permissible EM outcome $(E, A)$, let $f^{ToE}(E, A)=(E^{ToE}, A^{ToE})$ be the outcome produced by the ToE process. 
Then, every student $i \in I^U (E, A)$ is not Pareto improvable at $(E^{ToE}, A^{ToE})$. 
\end{proposition}

Proposition \ref{Prop_ToEmain} indicates that every student assigned to underdemanded majors at $(E, A)$ cannot achieve a Pareto improvement in any permissible outcome that Pareto dominates the ToE outcome $(E^{ToE}, A^{ToE})$.

We continue using Example \ref{Ex_incom} to show how the ToE process works.

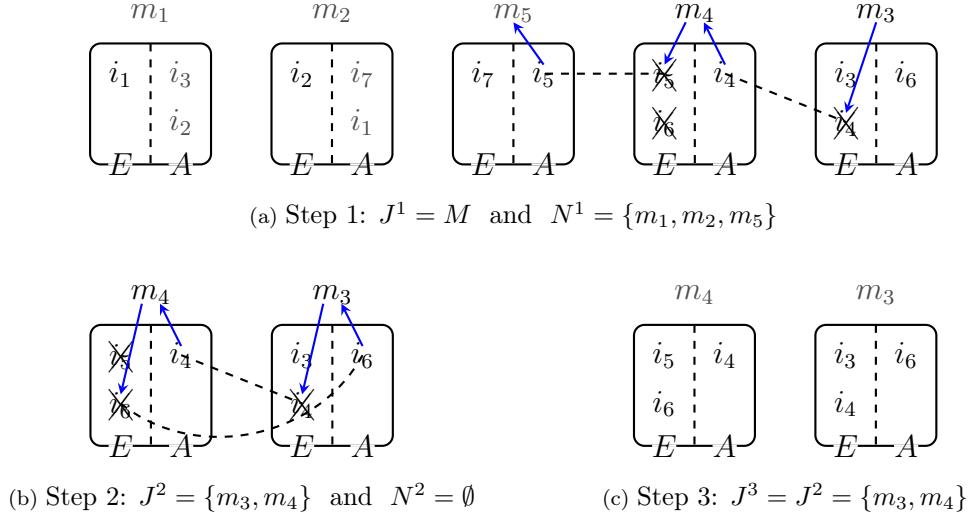
\begin{figure}[ht]
\begin{subfigure}[t]{\textwidth}
\centering
\begin{tikzpicture}[>=stealth, line width=0.8pt,scale=0.8]
    \foreach \i in {0, 1, 2, 3, 4} {
        \draw[rounded corners] ({-4 + \i * 3}, 1) -- ({-4 + \i * 3}, -1) -- ({-2 + \i * 3}, -1) -- ({-2 + \i * 3}, 1) -- cycle;
    }

    \draw[white, thick] (-3.7,-1) -- (-3.3,-1); 
    \draw[white, thick] (-2.7,-1) -- (-2.3,-1); 
    \draw[white, thick] (-0.7,-1) -- (-0.3,-1); 
    \draw[white, thick] (0.3,-1) -- (0.7,-1); 
    \draw[white, thick] (2.3,-1) -- (2.7,-1); 
    \draw[white, thick] (3.3,-1) -- (3.7,-1); 
    \draw[white, thick] (5.3,-1) -- (5.7,-1); 
    \draw[white, thick] (6.3,-1) -- (6.7,-1); 
    \draw[white, thick] (8.3,-1) -- (8.7,-1); 
    \draw[white, thick] (9.3,-1) -- (9.7,-1); 

    \node at (-3.5, -1) {$E$};
    \node at (-2.5,-1) {$A$};
    \node at (-0.5, -1) {$E$};
    \node at (0.5,-1) {$A$};
    \node at (2.5, -1) {$E$};
    \node at (3.5,-1) {$A$};
    \node at (5.5, -1) {$E$};
    \node at (6.5,-1) {$A$};
    \node at (8.5, -1) {$E$};
    \node at (9.5,-1) {$A$};

    \node at (-3,1.5) {$\textcolor{mycolor}{m_1}$};
    \node at (0,1.5) {$\textcolor{mycolor}{m_2}$};
    \node at (3,1.5) {$\textcolor{mycolor}{m_5}$};
    \node at (6,1.5) {$m_4$};
    \node at (9,1.5) {$m_3$};

    \node at (-3.5, 0.5) {$i_1$};
    \node at (-2.5, 0.5) {$\textcolor{mycolor}{i_3}$};
    \node at (-2.5, -0.3) {$\textcolor{mycolor}{i_2}$};
    \node at (-0.5, 0.5) {$i_2$};
    \node at (0.5, 0.5) {$\textcolor{mycolor}{i_7}$};
    \node at (0.5, -0.3) {$\textcolor{mycolor}{i_1}$};
    \node at (2.5, 0.5) {$i_7$};
    \node at (3.5, 0.5) {$i_5$};
    \node at (5.5, 0.5) {$\xcancel{i_5}$};
    \node at (5.5, -0.3) {$\xcancel{i_6}$};
    \node at (6.5, 0.5) {$i_4$};
    \node at (8.5, 0.5) {$i_3$};
    \node at (8.5,-0.3) {$\xcancel{i_4}$};
    \node at (9.5, 0.5) {$i_6$};
    \draw[dashed] (-3,1) to (-3,-1);
    \draw[dashed] (0,1) to (0,-1);
    \draw[dashed] (3,1) to (3,-1);
    \draw[dashed] (6,1) to (6,-1);
    \draw[dashed] (9,1) to (9,-1);
    \draw[dashed] (5.5, 0.5) to (3.5, 0.5);
    \draw[dashed] (6.5,0.5) to (8.5,-0.3);
    \draw[->, blue] (9,1.35) -- (8.5,-0.15);
    \draw[->, blue] (5.85,1.35) -- (5.5,0.65);
    \draw[->, blue] (3.5,0.65) -- (3,1.35);
    \draw[->, blue] (6.5,0.65) -- (6.15,1.35);
\end{tikzpicture}
\caption{\footnotesize{Step 1: $J^1=M$ \text{  and } $N^1=\{m_1, m_2, m_5\}$}}
\end{subfigure}

\vspace{0.5em}

\begin{center}

\begin{subfigure}[t]{0.40\textwidth}
\centering
\begin{tikzpicture}[>=stealth, line width=0.8pt,scale=0.8]
    \foreach \i in {3, 4} {
        \draw[rounded corners] ({-4 + \i * 3}, 1) -- ({-4 + \i * 3}, -1) -- ({-2 + \i * 3}, -1) -- ({-2 + \i * 3}, 1) -- cycle;
    }


    \draw[white, thick] (5.3,-1) -- (5.7,-1); 
    \draw[white, thick] (6.3,-1) -- (6.7,-1); 
    \draw[white, thick] (8.3,-1) -- (8.7,-1); 
    \draw[white, thick] (9.3,-1) -- (9.7,-1); 


    \node at (5.5, -1) {$E$};
    \node at (6.5,-1) {$A$};
    \node at (8.5, -1) {$E$};
    \node at (9.5,-1) {$A$};

    \node at (6,1.5) {$m_4$};
    \node at (9,1.5) {$m_3$};

    \node at (5.5, 0.5) {$\xcancel{i_5}$};
    \node at (5.5, -0.3) {$\xcancel{i_6}$};
    \node at (6.5, 0.5) {$i_4$};
    \node at (8.5, 0.5) {$i_3$};
    \node at (8.5,-0.3) {$\xcancel{i_4}$};
    \node at (9.5, 0.5) {$i_6$};

    \draw[dashed] (6,1) to (6,-1);
    \draw[dashed] (9,1) to (9,-1);
    \draw[dashed] (5.5,-0.3) to[bend right=50] (9.5,0.5);
    \draw[dashed] (6.5,0.5) to (8.5,-0.3);
    \draw[->, blue] (8.85,1.35) -- (8.5,-0.15);
    \draw[->, blue] (5.85,1.35) -- (5.5,-0.15);
    \draw[->, blue] (9.5,0.65) -- (9.15,1.35);
    \draw[->, blue] (6.5,0.65) -- (6.15,1.35);
\end{tikzpicture}
\caption{\footnotesize{Step 2: $J^2=\{m_3, m_4\}$ \text{  and } $N^2=\emptyset$}}
\end{subfigure}
\quad
\begin{subfigure}[t]{0.40\textwidth}
\centering
\begin{tikzpicture}[>=stealth, line width=0.8pt,scale=0.8]
    \foreach \i in {3, 4} {
        \draw[rounded corners] ({-4 + \i * 3}, 1) -- ({-4 + \i * 3}, -1) -- ({-2 + \i * 3}, -1) -- ({-2 + \i * 3}, 1) -- cycle;
    }


    \draw[white, thick] (5.3,-1) -- (5.7,-1); 
    \draw[white, thick] (6.3,-1) -- (6.7,-1); 
    \draw[white, thick] (8.3,-1) -- (8.7,-1); 
    \draw[white, thick] (9.3,-1) -- (9.7,-1); 


    \node at (5.5, -1) {$E$};
    \node at (6.5,-1) {$A$};
    \node at (8.5, -1) {$E$};
    \node at (9.5,-1) {$A$};

    \node at (6,1.5) {$\textcolor{mycolor}{m_4}$};
    \node at (9,1.5) {$\textcolor{mycolor}{m_3}$};

    \node at (5.5, 0.5) {$i_5$};
    \node at (5.5, -0.3) {$i_6$};
    \node at (6.5, 0.5) {$i_4$};
    \node at (8.5, 0.5) {$i_3$};
    \node at (8.5,-0.3) {$i_4$};
    \node at (9.5, 0.5) {$i_6$};

    \draw[dashed] (6,1) to (6,-1);
    \draw[dashed] (9,1) to (9,-1);
    
\end{tikzpicture}
\caption{\footnotesize{Step 3: $J^3=J^2=\{m_3, m_4\}$}}
\end{subfigure}
\caption{The ToE process in Example \ref{Ex_incom} given the permissible EM outcome $(E^7, A^7)$ as input}\label{Fig_Ex_ToE}
\end{center}
\end{figure}

Take the permissible EM outcome produced by the EM mechanism as an input, so that set $(E^1, A^1)=(\{i_1,i_2,i_3,i_7\}, \{i_4,i_5,i_6\})$ and $J^1=M$. 

At Step $1$, major $m_4$ points to $i_5$, and major $m_3$ points to $i_4$; $i_5$ points to the applied major $m_5$ and $i_4$ points to the applied major $m_4$. 
As shown in Figure \ref{Fig_Ex_ToE}(a), we have $N^1=\{m_1,m_2,m_5\} \neq \emptyset$ and hence go to process (a). 
Major $m_5$ evokes $i_5$'s transfer-in eligibility. 
Let $(E^2, A^2)=(\{i_1,i_2,i_3,i_7\}, \{i_4,i_6\})$, $J^2=J^1\setminus N^1=\{m_3,m_4\}$, and go to Step 2.

At Step $2$, major $m_3$ points to $i_4$, and major $m_4$ points to $i_6$; 
$i_4$ points to the applied major $m_4$, and $i_6$ points to the applied major $m_3$. 
As shown in Figure \ref{Fig_Ex_ToE}(b), we have $N^2=\emptyset$ and hence go to process (b). 
There exists a transfer-out exchangeable cycle $C^{\mathrm{out}}=(m_3, i_4, m_4, i_6, m_3)$. 
Hence, major $m_3$ gives transfer-out eligibility to $i_4$; 
major $m_4$ gives transfer-out eligibility to student $i_6$, and student $i_5$ who does not have transfer-in eligibility but has higher transfer-in priority than $i_6$.
Let $(E^3, A^3)=(\{i_1,i_2,i_3,i_4, i_5, i_6, i_7\}, \{i_4,i_6\})$, $J^3=J^2=\{m_3,m_4\}$, and go to Step 3.

At Step 3, neither major $m_3$ nor major $m_4$ have a transfer-out applicant without transfer-in eligibility. 
As shown in Figure \ref{Fig_Ex_TiE}(c), $N^3=\{m_1,m_2\}$ and hence go to process (a). 
Since $T^{in}_{m_3}=T^{in}_{m_4}=\emptyset$, set $(E^4, A^4)=(\{i_1,i_2,i_3,i_4, i_5, i_6, i_7\}, \{i_4,i_6\})$, $J^4=\emptyset$, and go to Step 4.

At Step $4$, since $J^4=\emptyset$, the process stops. 

In summary, the outcome produced by the ToE process is $f^{ToE}(E, A)=(E^\diamond, A^\diamond)=(\{i_1,i_2,i_3,i_4, i_5, i_6, i_7\}, \{i_4,i_6\})$ and this is a permissible and Pareto efficient outcome.

\subsection{Efficiency-adjusted EM Mechanisms}

In this section, we introduce two efficiency-adjusted EM mechanisms. 
These are based on both the TiE process and the ToE process. 
The main difference between the two mechanisms is the sequence of running the TiE process and the ToE process. 

We first show that both the TiE process and the ToE process are essential components of our proposed efficient mechanisms. 

\begin{lemma}\label{Le_eff01}
Given a permissible EM outcome $(E, A)$, the outcomes $f^{TiE}(E, A)$ and $f^{ToE}(E, A)$ may not be Pareto efficient. 
\end{lemma}

Lemma \ref{Le_eff01} indicates that applying only the TiE process or only the ToE process may be insufficient to find permissible and Pareto efficient outcomes. 

However, we will demonstrate that starting with a permissible EM outcome, running the TiE process and ToE process sequentially is sufficient to produce a Pareto efficient outcome. 
Moreover, the order of prioritizing the TiE process and the ToE process may result in different Pareto efficient outcomes.

\subsubsection{The TiE-prioritized efficiency-adjusted EM Mechanism} 

The first efficiency-adjusted EM mechanism prioritizes the TiE process. 
Given a permissible EM outcome as input, the EAEM-TiE mechanism first runs the TiE process, followed by the ToE process. 
A formal description of the mechanism is as follows.

\begin{center}
    \textbf{The TiE-prioritized efficiency-adjusted EM Mechanism $f^{E1}$ (EAEM-TiE)}
\end{center}
\begin{itemize}     
    
    \item[\textbf{Stage $1$}] Given a permissible EM outcome $(E, A)$, run the TiE process and obtain the outcome $(\hat{E}^{S1}, \hat{A}^{S1})$. 

    \item[\textbf{Stage $2$}] Given the outcome $(\hat{E}^{S1}, \hat{A}^{S1})$, run the ToE process and obtain the final outcome $(\hat{E}^{S2}, \hat{A}^{S2})$. 
   
\end{itemize}

Let $f^{E1} (E, A)=(\hat{E}^{S2}, \hat{A}^{S2})$ be the outcome produced by the mechanism $f^{E1}$. 
Let $K$ be the last step of the mechanism $f^{E1}$, and $(E^{k}, A^{k})$ be the outcome produced at step $k$. Note that step $k$ should be a step in either the TiE process or the ToE process. 

According to the mechanism $f^{E1}$, it is easy to obtain the following lemma. 

\begin{lemma}\label{Le_eff1}
Under the outcome $f^{E1} (E, A)=(\hat{E}^{S2}, \hat{A}^{S2})$, 
$\min_{\Pout_m} \hat{E}^{S2}_m \in \hat{A}^{S2} \cap \hat{E}^{S2}$ holds for each major $m \in M$. 
\end{lemma}

According to the process of mechanism $f^{E1}$, at step $1$, the input outcome $(E^{1}, A^{1})=(E, A)$ is a permissible EM outcome. 
Moreover, the outcome $(E^{k}, A^{k})$ produced in each step $k \geq 2$ of the mechanism $f^{E1}$ is a permissible outcome and (weakly) Pareto dominates the outcome $(E^{k-1}, A^{k-1})$ at step $k-1$. 
Formally, 

\begin{lemma}\label{Le_E1PD}
    The outcome $(E^{k}, A^{k})$ is permissible and $(E^{k} \cap A^{k}) \subseteq  (E^{k+1} \cap A^{k+1})$ holds true for each step $k \leq K$.
\end{lemma}

\begin{theorem}\label{Th_maineff}
The EAEM-TiE mechanism $f^{E1}$ is efficient. 
\end{theorem}

We continue using Example \ref{Ex_incom} to show the processes of the EAEM-TiE mechanism. 

In the Stage $1$, given the permissible EM outcome $(E, A)=(\{i_1,i_2,i_3,i_7\}, \{i_4,i_5,i_6\})$ produced by the EM mechanism, we run the TiE process and obtain the outcome $(\hat{E}^{S1}, \hat{A}^{S1})=f^{TiE}(E, A)=(E^*, A^*)=(\{i_1,i_2\}, \{i_1, i_2, i_3,i_4,i_5,i_6, i_7\})$ (see the analysis in section 5.2).

In the Stage $2$, given the outcome $(E^*, A^*)$, we run the ToE process and obtain the final outcome $f^{E1}(E, A)=(\hat{E}^{S2}, \hat{A}^{S2})=(E^*, A^*)$. 
We can see that no transfer-out exchangeable cycle is executed.  
The reason is that student $i_3$ is an interrupter of both the major swap between $i_1$ and $i_2$ and that between $i_4$ and $i_6$. 
Which swap can be executed depends on whether $i_3$ has transfer-in eligibility or transfer-out eligibility. 
Since the TiE process runs first in the EAEM-TiE mechanism, $i_3$ has transfer-in eligibility at outcome $(E^*, A^*)$, so that the major swap between $i_1$ and $i_2$ is executed via a transfer-in exchangeable cycle.

In summary, the EAEM-TiE mechanism $f^{E1}$ obtains the Pareto efficient outcome $f^{E1}(E, A)=(\hat{E}^{S2}, \hat{A}^{S2})=(E^*, A^*)$.

\subsubsection{The ToE-prioritized efficiency-adjusted EM Mechanism}

The second efficiency-adjusted EM mechanism prioritizes the ToE process. 
Given a permissible EM outcome as input, the mechanism first runs the ToE process, followed by the TiE process. 
A formal description of the mechanism is as follows.

\begin{center}
    \textbf{The ToE-prioritized efficiency-adjusted EM Mechanism $f^{E2}$ (EAEM-ToE)}
\end{center}
\begin{itemize}    
    
    \item[\textbf{Stage $1$}] Given a permissible EM outcome $(E, A)$, run the ToE process and obtain outcome $(\tilde{E}^{S1}, \tilde{A}^{S1})$. 

    \item[\textbf{Stage $2$}] Given the outcome $(\tilde{E}^{S1}, \tilde{A}^{S1})$, run the TiE process and obtain the final outcome $(\tilde{E}^{S2}, \tilde{A}^{S2})$. 
   
\end{itemize}

Let $f^{E2} (E, A)=(\tilde{E}^{S2}, \tilde{A}^{S2})$ be the outcome produced by the mechanism $f^{E1}$. 

\begin{theorem}\label{Th_maineff2}
The EAEM-ToE mechanism $f^{E2}$ is efficient. 
\end{theorem}

Since we can use the same logic to prove Theorem \ref{Th_maineff2}, we omit the proof.

We continue using Example \ref{Ex_incom} to show the processes of the EAEM-ToE mechanism. 

In Stage $1$, given the permissible EM outcome $(E, A)=(\{i_1,i_2,i_3,i_7\}, \{i_4,i_5,i_6\})$ produced by the EM mechanism, we run the ToE process and obtain the outcome $(\tilde{E}^{S1}, \tilde{A}^{S1})=f^{ToE}(E, A)=(E^\diamond, A^\diamond)=(\{i_1,i_2,i_3,i_4, i_5, i_6, i_7\}, \{i_4,i_6\})$ (see the analysis in section 5.3).

In Stage $2$, given the outcome $(E^\diamond, A^\diamond)$, we run the TiE process and obtain the final outcome $f^{E2}(E, A)=(\tilde{E}^{S2}, \tilde{A}^{S2})=(E^\diamond, A^\diamond)$. 
We can see that no transfer-in exchangeable cycle is executed. 
Since the ToE process runs first in the EAEM-ToE mechanism, $i_3$ has transfer-out eligibility at outcome $(E^\diamond, A^\diamond)$, so that the major swap between $i_4$ and $i_6$ is executed via a transfer-out exchangeable cycle. 

In summary, the EAEM-ToE mechanism $f^{E2}$ obtains the Pareto efficient outcome $f^{E2}(E, A)=(\tilde{E}^{S2}, \tilde{A}^{S2})=(E^\diamond, A^\diamond)$.

From Example \ref{Ex_incom}, we can see that the two efficient mechanisms $f^{E1}$ and $f^{E2}$ produce different Pareto efficient outcomes. 
Specifically, given the permissible EM outcome $(E, A)=(\{i_1,i_2,i_3,i_7\}, \{i_4,i_5,i_6\})$ as input, the EAEM-TiE mechanism executes a transfer-in exchangeable cycle $\Cin=(m_1, i_2, m_2, i_1, m_1)$, while the EAEM-ToE mechanism executes a transfer-out exchangeable cycle $\Cout=(m_4, i_6, m_3, i_4, m_4)$. 
In this case,  which exchangeable cycles can be executed depends on whether the TiE process or the ToE process runs first. 


\subsection{Discussion} 

\subsubsection{Major transition program under various types of distributional constraint} 

In this section, we discuss how the proposed mechanisms perform in some special cases. 

We have shown in Proposition \ref{prop_incom} that no major transition mechanism can be efficient and have eligibility maximization in a major transition program with both floor and ceiling constraints. 
Unfortunately, we further demonstrate that when a major transition program faces only one type of constraint --- either ceiling constraints or floor constraints --- the incompatibility result still holds.

\begin{proposition}\label{prop_incom2}
In the major transition program with ceiling constraints, floor constraints, or both, there is no major transition mechanism that can be efficient and have eligibility maximization. 
\end{proposition} 

First, we discuss how the proposed mechanisms perform in a major transition program, in which each major only faces a ceiling constraint (referred to as \textit{a major transition program with ceiling constraints}). 
In practice, to support students' transfer, some CMT programs impose almost no restrictions on transfer-out applications. 
However, applicants must obtain transfer-in eligibility from their applied majors to make their transfer successfully. 
In this case, each major has no floor constraint on the distribution of its students, i.e., $\underline{q}_m=0$ for each $m \in M$.

Consider any permissible EM outcome $(E, A)$. 
If major $m \in M$ does not distribute transfer-out eligibility to all transfer-out applicants, that is, $E_m \subset \omega_m$, major $m$ can permissibly expand its transfer-out eligibility set and transfer-in eligibility set by setting $E'_m=\omega_m$ and $A'=A$, which causes the contradiction that $(E, A)$ is a permissible EM outcome. 
Therefore, the following lemma holds. 

\begin{lemma}\label{Le_dis_01}
In a major transition program with ceiling constraints, every permissible EM outcome $(E, A)$ has $E_m=\omega_m$ for each major $m \in M$. 
\end{lemma}

According to Lemma \ref{Le_dis_01} and Lemma \ref{Le_AT}, in the major transition program with unbound floor constraints, there is no underdemanded major in any permissible EM outcome. 
Thus, any Pareto improvement over a permissible EM outcome occurs through transfer-in exchangeable cycles. 

\begin{proposition}\label{Prop_diss01}
In a major transition program with ceiling constraints, we have $f^{TiE}(E, A)=f^{E1}(E, A)=f^{E2}(E, A)$, where $(E, A)$ is a permissible EM outcome. 
\end{proposition} 

Proposition \ref{Prop_diss01} implies that when each major does not face a floor constraint, a straightforward way to find a Pareto efficient outcome is to start with any permissible EM outcome and run the TiE process. Under these distributional constraints, the outcomes produced by the TiE process and the two proposed efficient mechanisms are the same.

Second, we consider the major transition program in which each major faces a floor constraint but is ceiling-unbound (referred to as \textit{a major transition program with floor constraints}). 
When each major $m \in M$ does not face a ceiling constraint, i.e., $\overline{q}_m \geq \vert \omega_m \vert +\vert \alpha_m \vert$, any permissible EM outcome $(E, A)$ has $A_m=\alpha_m$ for each major $m \in M$. 
Formally, 

\begin{lemma}\label{Le_dis_02}
In a major transition program with floor constraints, every permissible EM outcome $(E, A)$ has $A_m=\alpha_m$ for each major $m \in M$. 
\end{lemma} 

This implies that there is no overdemanded major in any permissible EM outcome. 
Therefore, any Pareto improvement over a permissible EM outcome occurs through transfer-out exchangeable cycles. 
Formally, 

\begin{proposition}\label{Prop_diss02}
In a major transition program with floor constraints, we have $f^{ToE}(E, A)=f^{E1}(E, A)=f^{E2}(E, A)$, where $(E, A)$ is a permissible EM outcome. 
\end{proposition}

Proposition \ref{Prop_diss02} implies that when each major faces no ceiling constraint, a straightforward way to find a Pareto efficient outcome is to start with any permissible EM outcome and run the ToE process. In this kind of distributional constraint, the outcomes produced by the ToE process and the two proposed efficient mechanisms are the same.



\subsubsection{Exchange program under balanced conditions}

There is a growing literature studying the exchange markets under balanced conditions \citep{DU2019,C2022,KK2023}. 
In the context of major transition programs, the balanced condition refers to the situation where each major $m$ has $\vert \omega_m \vert =\overline{q}_m=\underline{q}_m$. 
Therefore, we have the following lemma.

\begin{lemma}\label{Le_diss_balance}
In the major transition program under the balanced condition, the outcomes $(I, \emptyset)$ and $(\emptyset, I)$ are permissible EM outcomes. 
\end{lemma}

To satisfy balanced conditions, if there is a student transferring out of their current major, there must be a new student transferring into the major. Therefore, every student who successfully transfers majors is involved in an exchange cycle. 
Based on Lemma \ref{Le_diss_balance}, we show two simple ways to find Pareto efficient outcomes under the balanced condition. 

The first way is to start with the permissible EM outcome $(I, \emptyset)$. 
Every student has transfer-out eligibility, so that any Pareto improvement over $(E, A)$ occurs through transfer-in exchangeable cycles. 
This indicates that, given the permissible EM outcome $(I, \emptyset)$, we can simply apply the TiE process to find a Pareto efficient outcome.

The second way is to start with the permissible EM outcome $(\emptyset, I)$.
Every student has transfer-in eligibility, so that any Pareto improvement over $(E', A')$ occurs through transfer-out exchangeable cycles. 
This implies that, given the permissible EM outcome $(\emptyset, I)$, we can simply apply the ToE process to find a Pareto efficient outcome. 
Formally, 

\begin{proposition}\label{Prop_diss03}
In a major transition program under the balanced condition, the following two statements hold:  

(1) $f^{TiE}(I, \emptyset)=f^{E1}(I, \emptyset)=f^{E2}(I, \emptyset)$; 

(2) $f^{ToE}(\emptyset, I)=f^{E1}(\emptyset, I)=f^{E2}(\emptyset, I)$.  

\end{proposition}

We can apply the same logic used in the proof of Proposition \ref{Prop_diss01}, so we will omit the proof here.

\section{Conclusion}

Major transitions play a vital role in supporting students as their academic interests and career goals evolve. 
By observing the inefficiencies in the main mechanism of the CMT program, in order to improve student welfare, we explore how to reassign students to majors in an efficient way, while respecting each major's dual priority and distributional constraints. 
Specifically, we introduce the concept of eligibility maximization and highlight its beneficial properties, and propose two efficient mechanisms based on two proposed processes for finding transfer-in and transfer-out exchangeable cycles, respectively. 
The efficient mechanisms produce Pareto efficient outcomes among those that  ensure a fair distribution of transfer-out and transfer-in eligibility among students. 

This paper provides a fresh framework for designing efficient mechanisms that account for both distributional and dual priority-respecting constraints. 
This framework can be extended to various exchange markets, such as interdistrict school choice, multi-center kidney exchange, tuition exchange programs, and job rotation programs, where fairness and distributional balance are crucial. 
Through this framework, we provide insights into achieving efficiency under distributional and dual priority-respecting constraints. 
An important direction for future research will be the design of efficient mechanisms when participants are allowed to apply for multiple positions. 
This would involve exploring the strategic problems faced by students, and how to accommodate multiple applications while ensuring fairness, efficiency, and adherence to any relevant constraints.


\bibliographystyle{plainnat}
\bibliography{MT_BIB}

\appendix

\section{Proofs}\label{App_proof}


\noindent \textbf{Proof of Proposition~\ref{Prop_EMP}}: By Proposition \ref{Prop_M-EM1}, we can directly obtain the result. 
\qed


\noindent \textbf{Proof of Lemma~\ref{Le_AT}}: 
To the contrary, suppose that in a permissible EM outcome $(E, A)$, there exists some major $m$ such that $\bar{E}_m \neq \emptyset$ and $\bar{A}_m \neq \emptyset $. 
Let $i_1 = \max_{\Pout_m} \bar{E}_m$ and $i_2 = \max_{\Pin_m} \bar{A}_m$. 
There are three possible cases. 

\textit{Case (1)}. If $i_1 \notin A_{-m}$. Then, $i_1$ remains in major $m$ even if major $m$ gives transfer-out eligibility to $i_1$. In this case, major $m$ can permissibly expand its transfer-out eligibility set and transfer-in eligibility set via $(E'_{m}, A'_{m})$, where $E'_m = E_m \cup \{i_1\}$ and $A'_m=A_m$, because the new outcome $(E', A')=(E'_m \cup E_{-m}, A)$ does not violate $m$'s dual priority and distributional constraints. 
This contradicts the fact that $(E, A)$ is a permissible EM outcome.

\textit{Case (2)}. If $i_2 \notin E_{-m}$. Then, $i_2$ remains in major $\omega_{i_2} \neq m$ even if major $m$ gives transfer-in eligibility to $i_2$. 
In this case, major $m$ can permissibly expand its transfer-out eligibility set and transfer-in eligibility set via $(E'_{m}, A'_{m})$, where $E'_m=E_m$ and $A'_m = A_m \cup \{i_2\}$, because the new outcome $(E', A')=(E, A'_m \cup A_{-m})$ does not violate $m$'s dual priority and distributional constraints. 
This contradicts the fact that $(E, A)$ is a permissible EM outcome.

\textit{Case (3)}. If $i_1 \in A_{-m}$ and $i_2 \in E_{-m}$. Then, major $m$ can permissibly expand its transfer-out eligibility set and transfer-in eligibility set via $(E'_{m}, A'_{m})$, where $E'_m = E_m \cup \{i_1\}$ and $A'_m = A_m \cup \{i_2\}$. 
The new outcome $(E',A') = (E'_m \cup E_{-m}, A'_m \cup A_{-m} )$ respects $m$'s distributional constraints because in the new outcome $(E', A')$, student $i_1$ transfers out from $m$ and student $i_2$ transfers into $m$ so that $\vert \mu_m (E', A') \vert = \vert \mu_m (E, A) \vert$ holds true. 
Moreover, the new outcome $(E',A')$ respects $m$'s dual priority because $i_1$ has the highest priority among transfer-out applicants who do not obtain transfer-out eligibility from $m$ and $i_2$ has the highest priority among transfer-in applicants who do not obtain transfer-in eligibility from $m$.  
This  contradicts the fact that $(E, A)$ is a permissible EM outcome. 
\qed


    



    
\noindent \textbf{Proof of Proposition~\ref{Prop_AT2}}: 
Given a permissible EM outcome $(E, A)$, let $(E',A')$ be a permissible outcome that Pareto dominates $(E, A)$. 

\underline{Proof of statement (1)}: 
We first show the ``Only if'' part. 
Suppose that there exists some student $i$ such that $\mu_i {(E,A) } \in M^O(E, A)$ but $\mu_i {(E',A')} \notin M^O(E, A)$. 
If $\alpha_i = \mu_i {(E,A) } \in M^O(E, A)$ and $\omega_i = \mu_i {(E',A') } \notin M^O(E, A)$, then $i$ is strictly worse off at $(E',A')$ than at $(E,A)$, which contradicts that $(E',A')$ Pareto dominates $(E,A)$.
Otherwise, we have $\omega_i = \mu_i {(E,A) } \in M^O(E, A)$ and $\alpha_i = \mu_i {(E',A') } \notin M^O(E, A)$. 
Then we have $i \in E$ (because major $\omega_i \in M^O(E, A)$ gives all its students transfer-out eligibility) and $i \in A$ (because major $\alpha_i \notin M^O(E, A)$ gives all its applicants transfer-in eligibility). 
Therefore, $i \in E \cap A$ holds, so that $\mu_i {(E,A) } = \alpha_i$, which contradicts $\mu_i {(E,A) } = \omega(i) $.

Now we show the ``If'' part. 
Suppose that there exists some student $i$ such that $\mu_i {(E,A) } \notin M^O(E, A)$ and $\mu_i {(E',A')} \in M^O(E, A)$. 
Since $\mu_i {(E,A) } \in M^O(E, A)$ implies $\mu_i {(E',A')} \in M^O(E, A)$ by the proof of ``Only if'' part, we have $\{ i \} \cup \bigl[ \bigcup_{m \in M^O(E, A)} \mu_m (E, A) \bigr] \subseteq \bigcup_{m \in M^O(E, A)} \mu_m (E', A')$, and
\[
    \sum_{m \in M^O(E, A)} \vert \mu_m (E', A') \vert \geq 1 + \sum_{m \in M^O(E, A)} \vert \mu_m (E, A) \vert = 1 + \sum_{m \in M^O(E, A)} \overline{q}_m. 
\]
There is at least one major $m \in M^O(E, A)$ that violates its ceiling constraint at $(E',A')$, which contradicts the fact that $(E',A')$ is permissible.

\underline{Proof of statement (2)}: Using logic similar to that in the proof of statement (1), we can show that $\mu_i {(E,A) } \in M^U(E, A)$ if and only if $\mu_i {(E',A')} \in M^U(E, A)$.

\underline{Proof of statement (3)}: 
As an implication of statements (1) and (2), we have $\mu_i {(E,A) } \in M^F(E, A)$ if and only if $\mu_i {(E',A')} \in M^F(E, A)$. 
We further show that $\mu_i {(E,A)} \in M^F(E, A)$ implies $\mu_i {(E',A')} = \mu_i {(E,A)}$.
Suppose that there exists some student $i^*$ such that $\mu_{i^*} {(E,A)} \in M^F(E, A)$ but $\mu_{i^*} {(E',A')} \neq \mu_{i^*} {(E,A)}$.
If $\mu_{i^*} {(E',A')} = \omega_{i^*} \neq \mu_{i^*} {(E,A)} = \alpha_{i^*} $, then student $i^*$ is strictly worse off at $(E',A')$. 
This causes the contradiction that $(E',A')$ Pareto dominates $(E, A)$. 
Otherwise, $\mu_{i^*} {(E',A')} = \alpha_{i^*} \neq \mu_{i^*} {(E,A)} = \omega_{i^*}$ holds. 
Then, we have $i^* \in E \cap A$ because majors $ \omega_{i^*}, \alpha_{i^*} \in M^F(E, A)$. (Each balanced major gives transfer-out eligibility to every transfer-out applicant and transfer-in eligibility to every transfer-in applicant.) 
This causes a contraindication that $\mu_{i^*} {(E,A)} = \omega_{i^*}$.

\underline{Proof of statement (4)}: 
Suppose that there exists some major $m^* \in M^O(E, A)$ such that $ \vert \mu_{m^*} {(E',A')} \vert \neq \vert \mu_{m^*} {(E,A)} \vert $, then it must be $\vert \mu_{m^*} {(E',A')} \vert < \vert \mu_{m^*} {(E,A)} \vert =  \overline{q}_{m^*}$. 
Then we have 
\[
    \sum_{m \in M^O(E, A)} \vert \mu_m (E', A') \vert < \sum_{m \in M^O(E, A)} \overline{q}_m = \sum_{m \in M^O(E, A)} \vert \mu_m (E, A) \vert
\]
There exists at least one student $i^*$ such that $\mu_{i^*} {(E,A) } \in M^O(E, A)$ and $\mu_{i^*} {(E',A')} \notin M^O(E, A)$. This contradicts statement (1).
Therefore, we have $\vert \mu_m (E, A) \vert = \vert \mu_m (E', A') \vert$ for each major $m \in M^O(E, A)$.

Suppose that there exists some major $m^* \in M^U(E, A)$ such that $ \vert \mu_{m^*} {(E',A')} \vert \neq \vert \mu_{m^*} {(E,A)} \vert $, then it must be $\vert \mu_{m^*} {(E',A')} \vert > \vert \mu_{m^*} {(E,A)} \vert =  \underline{q}_{m^*}$. 
Then we have 
\[
    \sum_{m \in M^U(E, A)} \vert \mu_m (E', A') \vert > \sum_{m \in M^U(E, A)} \underline{q}_m = \sum_{m \in M^U(E, A)} \vert \mu_m (E, A) \vert
\]
There exists at least one student $i^*$ such that $\mu_{i^*} {(E,A) } \notin M^U(E, A)$ and $\mu_{i^*} {(E',A')} \in M^U(E, A)$. This contradicts statement (2).
Therefore, we have $\vert \mu_m (E, A) \vert = \vert \mu_m (E', A') \vert$ for each major $m \in M^U(E, A)$.

Finally, by statement (3), $\vert \mu_m (E, A) \vert = \vert \mu_m (E', A')  \vert$ holds for each major $m \in M^F(E, A)$.
\qed



\noindent \textbf{Proof of Proposition~\ref{Prop_M-EM1}}: 
Let $K$ be the last step of the EM mechanism and hence the outcome produced by the EM mechanism is $(E^K, A^K)$. 

\underline{(1) The outcome $(E^K,A^K)$ respects each major's distributional constraints.}

First, we show that the outcome $(E^k, A^k)$ in each step $k \in \{1, \dots, K\}$ of the EM mechanism satisfies each major's ceiling constraints by induction. 
Consider any major $m \in M$.
For the base case, we have $\vert \mu_m {(E^1,A^1)} \vert = \vert \omega_m \vert \leq \overline{q}_m$. 
In any step $1 \leq k-1 < K$, suppose that we have $\vert \mu_m {(E^{k},A^{k})} \vert \leq \overline{q}_m$.
We will show $\vert \mu_m {(E^{k+1},A^{k+1})} \vert \leq \overline{q}_m$ holds true at step $k$. 
According to the EM mechanism, if at step $k$, major $m$ is not transfer-in expandable and its floor constraint is not violated, the statement is directly proved. 
Otherwise, there are two cases. 

\textit{Case (1)}. If at step $k$, major $m$ is transfer-in expandable and gives transfer-in eligibility to the student $i=\max_{\Pin_m} \bar{A}^{k}_m$. 
By the definition of the transfer-in expandable major, either the student $i$ has no transfer-out eligibility or the major $m$ has vacant position(s). 
In the first case, the student $i$ does not transfer in $m$ at $(E^{k+1}, A^{k+1})$ and hence $\vert \mu_m {(E^{k+1},A^{k+1})} \vert=\vert \mu_m {(E^{k},A^{k})} \vert$ holds true. 
In the second case, $\vert \mu_m {(E^{k+1},A^{k+1})} \vert=\vert \mu_m {(E^{k},A^{k})} \vert+1 \leq \overline{q}_m$ holds true.

\textit{Case (2)}. If at step $k$, major $m$'s floor constraint is violated.
Then, we know that $\vert \mu_m {(E^{k},A^{k})} \vert < \underline{q}_m \leq \overline{q}_m$. 
Major $m$ revokes the transfer-out eligibility of the student $i'=\min_{\Pout_m}E^k_m$. 
If $i' \notin A^k$, then $\vert \mu_m {(E^{k+1},A^{k+1})} \vert=\vert \mu_m {(E^{k},A^{k})} \vert < \overline{q}_m$ holds true. 
Otherwise, $i' \in A^k$, then $\vert \mu_m {(E^{k+1},A^{k+1})} \vert=\vert \mu_m {(E^{k},A^{k})} \vert+1 \leq \overline{q}_m$ holds true. 

By induction, the EM outcome $(E^K,A^K)$ respects each major's ceiling constraint. 

Second, we show that the outcome $(E^K,A^K)$ respects each major's floor constraint.
Suppose that there exists some major $m \in M$ such that $\vert \mu_m {(E^K,A^K)} \vert < \underline{q}_m$. 
According to the EM mechanism, if $E^K_m \neq \emptyset$ at step $K$, then step $K$ is not the last step. 
Therefore, $E^K_m = \emptyset$ so that all current students in major $m$ do not have transfer-out eligibility. 
This implies that every current student in major $m$ fails to transfer their major, and hence $\vert \mu_m {(E^K,A^K)} \vert \geq \underline{q}_m$ holds true. 

In summary, the outcome $(E^K,A^K)$ satisfies the ceiling and floor constraints of each major.

\underline{(2) The outcome $(E^K,A^K)$ respects each major's dual priority.} 

The EM mechanism begins with a permissible outcome $(E^1, A^1)=(I, \emptyset)$. 
According to the EM mechanism, at each step $k$, either $E^{k+1} \subset E^k$ or $A^{k} \subset A^{k+1}$, but not both. 
If $E^{k+1} \subset E^k$ holds, then there exists at least one major that revokes the transfer-out eligibility of the lowest priority student among its transfer-out applicants who have transfer-out eligibility at $(E^k, A^k)$. By remark~\ref{Remark_EM}, every student whose transfer-out eligibility is revoked by their initial major at some step will not obtain transfer-out eligibility at later steps. 
Therefore, the EM outcome $(E^K, A^K)$ respects each major's transfer-out priority. 
Similarly, 
if $A^{k} \subset A^{k+1}$ holds, then there exists at least one transfer-in expandable major that gives transfer-in eligibility to the highest priority student among those transfer-in applicants who do not have transfer-in eligibility at $(E^k, A^k)$. 
By remark~\ref{Remark_EM}, every student who obtains transfer-in eligibility at some step will not lose transfer-in eligibility at later steps. 
Therefore, the EM outcome $(E^K, A^K)$ respects each major's transfer-in priority. 

\underline{(3) The outcome $(E^K, A^K)$ has eligibility maximization.}

We first show that the following claim holds. 

\noindent\textbf{Claim 1}: In each step $k \in \{1,2, \dots, K\}$ of the EM mechanism and for each major $m \in M$, at least one of the two equations holds: (1) $E^k_m = \omega_m$ and (2) $A^k_m = \alpha_m$. 
\proof 
The EM mechanism starts from the permissible outcome $(E^1, A^1)=(I, \emptyset)$. 
If any major $m$ has never revoked its transfer-out eligibility in the process of the EM mechanism, then we have $E^k_m = \omega_m$ for all $k \in \{1, \dots, K\}$.
Otherwise, let step $k$ be the step in which major $m$ revokes its transfer-out eligibility for the first time. That is, $E^\kappa_m = \omega_m$ for all $\kappa \in \{1,2,\dots, k\}$ and $E^{k+1}_m \neq \omega_m$.
Therefore, at step $k$, major $m$'s floor constraint is violated, such that $\vert \mu^k_m \vert < \underline{q}_m \leq \overline{q}_m$. 
If $A^k_m \neq \alpha_m$, then major $m$ is transfer-in expandable. According to the EM mechanism, no major will evoke their transfer-out eligibility at step $k$, which causes a contradiction.  
Hence, we have $A^k_m = \alpha_m$. 
By Remark~\ref{Remark_EM}, we have $A^\kappa_m = \alpha_m$ for all $\kappa \in \{k, k+1, \dots, K\}$. 
\qed

Now, suppose that the outcome $(E^K, A^K)$ does not have eligibility maximization. 
That is, there exists some major $m \in M$ that can permissibly expand its transfer-out eligibility set and transfer-in eligibility set at the permissible outcome $(E^K, A^K)$. 
By definition, there exist $E'_m \supseteq E_m$ and $A'_m \supseteq A_m$ with at least one proper superset, such that the new outcome $(E',A') = (E'_m \cup E_{-m}, A'_m \cup A_{-m} )$ respects $m$'s dual priority and distributional constraints.
There are two cases. 

\textit{Case (1)}. $A^K_m \subsetneq A'_m \subseteq \alpha_m$. This implies that $\alpha_m \setminus A^K_m \neq \emptyset$ and by Claim $1$, $\omega_m = E^K_m = E'_m$. 
Let $i^* = \max_{\Pin_m}\bar{A}^K_m$. 
By the definition of permissible expansion, we have $(A^K \cup \{i^*\} ) \subseteq A'_m$.
Since step $K$ is the last step of the EM mechanism, major $m$ is not transfer-in expandable at $(E^K, A^K)$. 
This implies that $i^* \in E$ and $\vert \mu_{(E^K,A^K)} \vert  = \overline{q}_m$ hold true. 
Since $E' = E^K$ and $A' \supseteq A^K \cup \{i^*\}$ and $\mu_m (E', A') \supseteq \mu_m {(E^K,A^K)} \cup \{i^*\}$, major $m$'s ceiling constraint is violated at $(E', A')$, which causes a contradiction. 

\textit{Case (2)}. $E^K_m \subsetneq E'_m \subseteq \omega_m$. 
This implies that $\omega_m \setminus E^K_m \neq \emptyset$ and by Claim $1$, $A'_m = A^K_m = \alpha_m$. 
Therefore, major $m$ has revoked the transfer-out eligibility of some of its transfer-out applicants in the EM mechanism. 
Let $i^* = \max_{\Pout_m} \bar{E}^K_m$ and $k$ be the latest step in which major $m$ revokes the transfer-out eligibility of student $i^*$. 
By Claim $1$, we have $E^{k}_m =  E^{k+1}_m \cup \{i^*\} = E^{K}_m \cup \{i^*\}$, $A^k_m = \alpha_m$, $\vert \mu_m {(E^{k}, A^{k})} \vert = \underline{q}_m - 1$ and $\vert \mu_m {(E^{k+1}, A^{k+1})} \vert = \underline{q}_m$. 
The latter two conditions imply that $i^* \in A^k_{-m}$, and by Remark \ref{Remark_EM}, we have $A^k_{-m} \subseteq A^K_{-m} = A'_{-m}$ and $E^k_{-m} \supseteq E^K_{-m}$.
By the definition of permissible expansion, $E^k_m = (E^K_m \cup \{i^*\} ) \subseteq E'_m$ holds.
Therefore, we have $E^k_m \cap A^k_{-m} \subseteq E'_m \cap A'_{-m}$ and $A^k_m \cap E^k_{-m} \supseteq A'_{m} \cap E'_{-m}$.
Recall that $\mu_m{(E',A')} = [ \omega_m \setminus (E'_m \cap A'_{-m})] \cup [A'_m \cap E'_{-m}] $ and $\mu_m {(E^{k},A^{k})} = [ \omega_m \setminus (E^{k}_m \cap A^{k}_{-m}] \cup [A^{k}_{m} \cap E^{k}_{-m}] $, we have $\mu_m (E', A') \subseteq \mu_m {(E^{k},A^{k})}$.
Given that $\vert \mu_m {(E^{k}, A^{k})} \vert = \underline{q}_m - 1$, the new outcome $(E', A')$ violates $m$'s floor constraint, which causes a contradiction. 
\qed


\noindent \textbf{Proof of Proposition~\ref{Prop_TiEc}}: 
We first show that the outcome $(E', A')$, where $E'=E$ and $A'=A \cup \bigcup_{m_l \in \Cin} \{i \in \bar{A}_{m_l} \vert i \Pineq_{m_l} i_{l+1}\}$ ($l \in \{0,...,L-1\}$), respects each major's dual priority. 
Since outcome $(E, A)$ is permissible and $E'=E$, the outcome $(E', A')$ respects every major's transfer-out priority. 
Since for each major $m \notin \Cin$, we have $A'_m=A_m$, so that the outcome $(E', A')$ respects the transfer-in priority of each major $m \in M \setminus \Cin$.
For each major $m \in \Cin$, every transfer-in applicant who does not obtain transfer-in eligibility at $(E, A)$ and has a higher priority than the applicant pointed by $m$ receives transfer-in eligibility from $m$ so that the outcome $(E', A')$ also respects $m$'s transfer-in priority. In summary, the outcome $(E', A')$ respects each major's dual priority. 

Next, we show that the outcome $(E', A')$ respects the distributional constraints of each major. 
Since for each major $m \in M \setminus \Cin$, we have $E'_m=E_m$ and $A'_m=A_m$, so that the outcome $(E',A')$ respects the distributional constraints of each major $m \in M \setminus \Cin$. 
According to the pointing process, each major $m \in \Cin$ points to the highest priority transfer-in applicant $i^*$ who has only transfer-out eligibility. 
This indicates that any transfer-in applicant $i \in \{i \in \bar{A}_m \vert i \Pin_{m} i^*\}$ has no transfer-out eligibility. 
Therefore, in each major $m \in \Cin$, one student transfers out and one student transfers in, so that $\vert \mu_m (E, A) \vert=\vert \mu_m {(E', A')} \vert$ holds true.
Since the outcome $(E,A)$ respects each major's distributional constraints, the outcome $(E', A')$ also respects each major's distributional constraints. 
\qed



\noindent \textbf{Proof of Proposition~\ref{Prop_ToEc}}: 
We first show that the outcome $(E', A')$, where $E'=E \cup \bigcup_{m_l \in \Cout} \{i \in \bar{E}_{m_l} \vert i \Pouteq_{m_l} i_{l}\}$ ($l \in \{1,...,L\}$) and $A'=A$, respects each major's dual priority. 
Since outcome $(E,A)$ is permissible and $A'=A$, the outcome $(E',A')$ respects every major's transfer-in priority. 
Since for each major $m \notin \Cout$, we have $E'_m=E_m$, so that the outcome $(E', A')$ respects the transfer-out priority of each major $m \in M \setminus \Cout$. 
For each major $m \in \Cout$, every transfer-out applicant who does not obtain transfer-out eligibility at $(E,A)$ and has a higher priority than the applicant pointed by $m$, obtains transfer-out eligibility from $m$ so that the outcome $(E',A')$ also respects $m$'s transfer-out priority. In summary, the outcome $(E',A')$ respects each major's dual priority.

We next show that the outcome $(E',A')$ respects the distributional constraints. Since for each major $m \in M \setminus \Cout$, we have $E'_m=E_m$ and $A'_m=A_m$ so that the outcome $(E',A')$ respects the distributional constraints of each major $m \in M \setminus \Cout$. According to the pointing process, each major $m \in \Cout$ points to the highest priority transfer-out applicant $i^*$ who only has transfer-out eligibility. This indicates that any transfer-out applicant $i \in \{i \in \bar{E}_m \vert i\Pout_{m} i^*\}$ has no transfer-in eligibility. 
Therefore, in each major $m \in \Cout$, one student transfers out and one student transfers in, so that $\vert \mu_m (E, A) \vert=\vert \mu_m {(E', A')} \vert$ holds true.
Since the outcome $(E,A)$ respects each major's distributional constraints, the outcome $(E', A')$ also respects each major's distributional constraints. 
\qed



\noindent \textbf{Proof of Lemma~\ref{Le_TiE}}: 
Suppose that there exists some student $i \in E \cap A$ who becomes worse off at outcome $f^{TiE}=(E^{TiE}, A^{TiE})$. 
By Remark~\ref{Rm_ParetoImprove}(b), $i \in A^{TiE}$ holds true so that $i \notin E^{TiE}$. 
According to the TiE process, the major $\omega_i$ at most evokes students who have lower transfer-out priority than $i^* = \min_{\Pout_{\omega_i}} E_{\omega_i} \cap A$. 
Since $i \in E \cap A$, we have $i \in E^{TiE}$, which contradicts the assumption. 
\qed


\noindent \textbf{Proof of Proposition~\ref{Prop_TiEmain}}: 
Suppose that there exists a permissible outcome $(E',A')$ Pareto dominating the TiE outcome $(E^{TiE}, A^{TiE})$ and $I' \cap I^O (E,A) \neq \emptyset$, where $I' = (E' \cap A') \setminus (E^{TiE} \cap A^{TiE})$ is the set of students that is strictly improved. 
Let $I^* = I' \cap I^O (E,A)$, and $M^* = \bigcup_{i \in I^*} \{ \omega_{i} \}$. 
By Lemma~\ref{Le_TiE}, $(E^{TiE}, A^{TiE})$ Pareto dominates the EM outcome $(E, A)$. Therefore, $(E',A')$ which Pareto dominates $(E^{TiE}, A^{TiE})$ also Pareto dominates $(E, A)$.  
This implies that for each $i \in I^*$, $\mu_i(E, A)=\omega_i$ holds, and hence each major $m \in M^*$ is overdemanded at $(E, A)$. 

Consider the major $m_1 \in M^*$ that is removed no later than any other majors in $M^*$ during the TiE process, and let $k$ be the step at which $m_1$ is removed.
Formally, we have $M^* \subseteq J^k$ and $m_1 \in N^k$.
Let $i_1 \in I^* \cap \omega_{m_1}$ be a student who is strictly improved at $(E', A')$ and whose initial major is $m_1$.
According to statement (4) of Proposition~\ref{Prop_AT2}, we have $\vert \mu_{m_1} (E', A') \vert = \vert \mu_{m_1} (E^{TiE}, A^{TiE}) \vert = \vert \mu_{m_1} (E, A) \vert$. 
Consequently, when there are students like $i_1$ who transfer out of major $m_1$ at $(E',A')$, there have to be an equal number of students transferring to major $m_1$ who become strictly better off at $(E', A')$ than at $(E^{TiE}, A^{TiE})$ . 
According to statement (1) of Proposition~\ref{Prop_AT2}, there exists a student $i_2 \in I^* \cap \alpha_{m_1}$. Let $m_2 = \omega_{i_2}$. 
Then $m_2 \in M^* \subseteq J^k$. 

At the beginning of step $k$ of the TiE process, since major $m_2$ is overdemanded at $(E, A)$, major $m_2$ has $E^k_{m_2} = \omega_{m_2} \ni i_2$. 
Moreover, $\mu_{i_2} (E^{TiE},A^{TiE}) = \omega_{i_2}=m_2$ implies that $i_2 \notin A^k_{m_1}$. So that $\bar{A}^k_{m_1} \cap E^k$ is not empty.
This implies that $m_1$ points to some student at step $k$. 
Since $m_1$ is removed from $J^k$ at this step, $m_1$ must point to an applicant $i_3 = \max_{\Pin_{m_1}} (\bar{A}^k_{m_1} \cap E^k)$ such that $m_3 \equiv \omega_{i_3} \notin J^k$ and $i_3 \Pin_{m_1} i_2$. 
According to the definition of the TiE process, we have $\mu_{i_3} (E^{TiE}, A^{TiE}) = m_3$.
Note that major $m_3 \notin J^k$ has evoked all of its redundant transfer-out eligibility (if any) at an earlier step $\kappa < k$ of the TiE process, and $i_3 \in E^k$ keeps her transfer-out eligibility, so it must be that $i_3 \Pout_{m_3} i^{*}$ where $i^{*} = \min_{\Pout_{m_3}} (E^\kappa_{m_3} \cap A^\kappa)$ is the lowest priority student who successfully transfers out from major $m_3$ at $(E^{TiE}, A^{TiE})$. We also have $i^* \in E' \cap A'$ since $i^*$ cannot be worse off at the Pareto dominant outcome $(E',A')$.
Now, $i^{*} \in E'$ implies $i_3 \in E'$; similarly, $i_2 \in A'$ implies $i_3 \in A'$. So $i_3 \in E' \cap A'$ is improved at $(E',A')$. 
However, $m_3 \notin J^k$ and $M^* \subseteq J^k$ imply $m_3 \notin M^*$. So,  $i_3 \in \omega_{m_3}$ should not be improved at $(E',A')$, which causes a contradiction. 
\qed



\noindent \textbf{Proof of Lemma~\ref{Le_ToE}}: 
Suppose that there exists some student $i \in E \cap A$ who becomes worse off at outcome $f^{ToE}=(E^{ToE}, A^{ToE})$. 
By Remark~\ref{Rm_ParetoImprove2}(b), $i \in E^{ToE}$ holds true, so that $i \notin A^{ToE}$. 
According to the ToE process, major $\alpha_i$ at most evokes the transfer-in eligibility of students who have lower transfer-out priority than $i^* = \min_{\Pin_{\alpha_i}} A_{\alpha_i} \cap E$. 
Since $i \in E \cap A$, we have $i \in A^{ToE}$, which causes a contradiction. 
\qed



\noindent \textbf{Proof of Proposition~\ref{Prop_ToEmain}}:
Suppose that there exists a permissible outcome $(E',A')$ Pareto dominating the ToE outcome $(E^{ToE}, A^{ToE})$ and $I' \cap I^U (E,A) \neq \emptyset$, where $I' = (E' \cap A') \setminus (E^{ToE} \cap A^{ToE})$ is the set of students that are strictly improved. 
Let $I^* = I' \cap I^U (E,A)$, and $M^* = \bigcup_{i \in I^*} \{ \alpha_{i} \}$. 
By Lemma~\ref{Le_ToE}, $(E^{ToE}, A^{ToE})$ Pareto dominates the EM outcome $(E, A)$. Therefore, $(E',A')$ which Pareto dominates $(E^{ToE}, A^{ToE})$ also Pareto dominates $(E, A)$. 
This implies that for each $i \in I^*$, $\mu_i(E, A)=\omega_i$ holds true, and hence each major $m \in M^*$ is underdemanded at $(E, A)$. 

Consider the major $m_1 \in M^*$ that is removed no later than any other majors in $M^*$ during the ToE process, and let $k$ be the step at which $m_1$ is removed.
Formally, we have $M^* \subseteq J^k$ and $m_1 \in N^k$.
Let $i_1 \in I^* \cap \alpha_{m_1}$ be a student who is strictly improved at $(E', A')$ and whose applied major is $m_1$.
According to the statement (4) of Proposition~\ref{Prop_AT2}, we have $\vert \mu_{m_1} (E', A') \vert = \vert \mu_{m_1} (E^{ToE}, A^{ToE}) \vert = \vert \mu_{m_1} (E, A) \vert$. 
Consequently, when there are students like $i_1$ who transfer in major $m_1$ at $(E',A')$, there has to be an equal number of students who transfer out of major $m_1$. 
According to statement (1) of Proposition~\ref{Prop_AT2}, there exists a student $i_2 \in I^* \cap \omega_{m_1}$. 
Let $m_2 = \alpha_{i_2}$. Then $m_2 \in M^* \subseteq J^k$.

At the beginning of step $k$ of the ToE process, since major $m_2$ is underdemanded at $(E, A)$, major $m_2$ has $A^k_{m_2} = \alpha_{m_2} \ni i_2$. 
Moreover, by Remark~\ref{Rm_ParetoImprove2}, $\mu_{i_2} (E^{ToE},A^{ToE}) = \omega_{i_2} $ implies that $i_2 \notin E^k_{m_1}$. So, $\bar{E}^k_{m_1} \cap A^k$ is not empty.
This implies that major $m_1$ points to some student at step $k$. Since $m_1$ is removed from $J^k$ at this step, $m_1$ must point to an applicant $i_3 = \max_{\Pout_{m_1}} (\bar{E}^k_{m_1} \cap A^k)$ such that $m_3 \equiv \alpha_{i_3} \notin J^k$ and $i_3 \Pout_{m_1} i_2$. According to the definition of the ToE process, we have $\mu_{i_3} (E^{ToE}, A^{ToE}) = m_2$.
Note that major $m_3 \notin J^k$ has evoked all of its redundant transfer-in eligibility (if any) at an earlier step $\kappa < k$ of the ToE process.
That $i_3 \in A^k$ implies that $i_3 \Pin_{m_3} i^{*}$ where $i^{*} = \min_{\Pin_{m_3}} (E^\kappa_{m_3} \cap A^\kappa)$ is the lowest priority student who successfully transfers in major $m_3$ at $(E^{ToE}, A^{ToE})$. We also have $i^* \in E' \cap A'$ since $i^*$ cannot be worse off at the Pareto dominant outcome $(E',A')$.
Now, $i^{*} \in A'$ implies $i_3 \in A'$; similarly, $i_2 \in E'$ implies $i_3 \in E'$. So $i_3 \in E' \cap A'$ is improved at $(E',A')$. 
However, $m_3 \notin J^k$ and $M^* \subseteq J^k$ imply that $m_3 \notin M^*$. So $i_3 \in \alpha_{m_3}$ should not be improved at $(E',A')$, which causes a contradiction.
\qed



\noindent \textbf{Proof of Lemma~\ref{Le_eff01}}:
We use an example to prove this. There are five majors $M=\{m_1,m_2,...,m_5\}$ and six students $I=\{i_1,i_2,...,i_6\}$. The distributional constraints, transfer-out priority, and transfer-in priority of each major are shown in the following table.

\begin{center}
\begin{tabular}{|c|c|l|l|}
\hline
    Major & $(\underline{q}_m, \overline{q}_m)$ & transfer-out priorities & transfer-in priorities \\\hline
    $m_1$ & $(0,1)$  & $\Pout_{m_1}: i_1$ & $\Pin_{m_1}: i_3, i_2$ \\\hline
    $m_2$ & $(0,1)$  & $\Pout_{m_2}: i_2$ & $\Pin_{m_2}:  i_1$ \\\hline
    $m_3$ & $(2,2)$  & $\Pout_{m_3}: i_4, i_3 $  & $\Pin_{m_3}: i_6$ \\\hline
    $m_4$ & $(2,2)$  & $\Pout_{m_4}: i_5, i_6$  & $\Pin_{m_4}: i_4$ \\\hline
    $m_5$ & $(0,1)$  & $\Pout_{m_5}: - $  & $\Pin_{m_5}: i_5$ \\\hline
\end{tabular}
\end{center} 

Given the permissible EM outcome $(E,A)=(\{i_1,i_2\},\{i_3,i_4,i_5,i_6\})$, the outcome produced by the TiE process is $f^{TiE}(E, A)=(\{i_1,i_2\},\{i_1,i_2, i_3, i_4,i_5,i_6\})$. 
The outcome $f^{TiE}(E, A)$ is not Pareto efficient because it can be Pareto dominated by the Pareto efficient outcome $(E^*,A^*)=(\{i_1,i_2,i_4,i_5,i_6\}, \{i_1,i_2,i_3,i_4,i_6\})$. 
Similarly, given the permissible EM outcome $(E,A)$, the outcome produced by the ToE process is $f^{ToE}(E, A)=(\{i_1,i_2,i_4,i_5,i_6\},\{i_3,i_4,i_6\})$.
The outcome $f^{ToE}(E, A)$ is not Pareto efficient because it can be Pareto dominated by the Pareto efficient outcome $(E^*,A^*)$. 
\qed



\noindent \textbf{Proof of Lemma~\ref{Le_E1PD}}:
Consider some step $k$ of the TiE process in stage 1. Suppose $i\in E^k\cap A^k$. By Remark~\ref{Rm_ParetoImprove}, we have $E^{k+1}\subseteq E^k$ and $A^k\subset A^{k+1}$. Then $i\in A^{k+1}$. By the rule of the TiE process, only the transfer-out eligibility of students who have transfer-out eligibility but no transfer-in eligibility can be revoked, which implies $i\in E^{k+1}$ and thus $i\in E^{k+1}\cap A^{k+1}$. Therefore, we have $(E^{k} \cap A^{k}) \subseteq  (E^{k+1} \cap A^{k+1})$. 

Similarly, consider some step $k$ of the ToE process in stage 2. Suppose $i\in E^k\cap A^k$. By Remark~\ref{Rm_ParetoImprove2}, we have $E^{k}\subseteq E^{k+1}$ and $A^{k+1}\subseteq A^k$. Then $i\in E^{k+1}$. By the rule of the ToE process, only the transfer-in eligibility of students who have transfer-in eligibility but no transfer-out eligibility can be withdrawn, which implies $i\in A^{k+1}$ and thus $i\in E^{k+1}\cap A^{k+1}$. Therefore, we have $(E^{k} \cap A^{k}) \subseteq  (E^{k+1} \cap A^{k+1})$.
\qed


\noindent \textbf{Proof of Theorem~\ref{Th_maineff}}:
Suppose that the mechanism $f^{E1}$ is not efficient. 
That is, given some permissible EM outcome $(E, A)$, the outcome $f^{E1}(E, A)=(\hat{E}^{S2}, \hat{A}^{S2})$ can be Pareto dominated by a permissible outcome $(E', A')$. 
To prove the theorem, we will use the following notations. 
\[
\tilde{I}=(E' \cap A') \setminus (\hat{E}^{S2} \cap \hat{A}^{S2}), \quad \mbox{ and }
\tilde{M}= \bigcup_{i \in \tilde{I}} \{\omega_i\}.
\]
Since $(E', A')$ Pareto dominates $(\hat{E}^{S2}, \hat{A}^{S2})$, which Pareto dominates $(\hat{E}^{S1}, \hat{A}^{S1})$ and $(E, A)$, then $(E', A')$ Pareto dominates $(\hat{E}^{S1}, \hat{A}^{S1})$ and $(E, A)$. 
By Proposition \ref{Prop_TiEmain}, $\tilde{I} \cap I^O (E, A) = \emptyset$ holds, so that we have $\tilde{I} \subseteq I^U (E, A)$ and $\tilde{M} \subseteq M^U (E, A)$.
By the definition of $M^U (E,A)$, we have $A_m = \alpha_m$ for each $m \in M^U (E,A)$. 
According to Remark~\ref{Rm_ParetoImprove}(b), we have $\hat{A}^{S1}_{m} = \alpha_m$ for every $m \in M^U (E, A)$.

Now consider stage $2$ (the ToE process) of mechanism $f^{E1}$. Given the outcome $(\hat{E}^{S1}, \hat{A}^{S1})$ produced by stage $1$ of the mechanism $f^{E1}$, let $(E^k, A^k)$ be the outcome produced by step $k$ of the ToE process in stage $2$ of $f^{E1}$, where $(E^1, A^1)=(\hat{E}^{S1}, \hat{A}^{S1})$ and $j^1=M$. 
Let $K$ be the last step of the ToE process. 
According to the mechanism $f^{E1}$, we know that $(E^K, A^K)=(\hat{E}^{S2}, \hat{A}^{S2})$. 
Then, we have $A^{k}_{m} = \alpha_m$ for every $m \in M^U (E, A)$ as long as major $m \in J^k$.

We will show $\tilde{M} = \emptyset$. 
Suppose not. 
Let $k$ be the first step in Stage 2 (the ToE process) of the mechanism $f^{E1}$, in which some major $m_1 \in \tilde{M}$ is excluded from the remaining major set $J^k$. That is $\tilde{M} \subseteq J^k$ and $m_1 \in N^k$.
Let $i_1 \in \tilde{I} \cap \omega_{m_1}$ be the student who is strictly improved in $\omega_{m_1}$.
As an implication of statement (4) of Proposition~\ref{Prop_AT2}, we know that $m_2 = \alpha_{i_1} \in \tilde{M} \subseteq J^k$. So we have $i_1 \in A^k_{m_2}$. 
If $i_1 \in E^k_{m_1}$, then we have $i_1 \in (A^k \cap E^k) \subseteq (\hat{A}^{S2} \cap \hat{E}^{S2})$, which contradicts the fact that $i_1$ is improved by $(E',A')$. So $i_1 \notin E^k_{m_1}$.
Now we know that $i_1 \in \bar{E}^k_{m_1} \cap A^k$, and thus $\bar{E}^k_{m_1} \cap A^k$ is not empty.
At step $k$ of the ToE process, major $m_1$ has a student to point to, but it is removed from $J^k$, so it must point to another student $i_2 = \max_{\Pout_{m_1}} [\bar{E}^k_{m_1} \cap A^k]$ such that $m_3 = \alpha_{i_2} \notin J^k$ and $i_2 \Pout_{m_1} i_1$.

According to the definition of the ToE process, we have $\mu_{i_2} (\hat{E}^{S2}, \hat{A}^{S2}) = m_1$.
Note that major $m_3 \notin J^k$ has withdrawn all of its redundant transfer-in eligibility (if any) at an earlier step $\kappa < k$ of the ToE process, and $i_2 \in A^k$ keeps her transfer-in eligibility, so it must be that $i_2 \Pin_{m_3} i_3$ where $i_3 = \min_{\Pin_{m_3}} (E^\kappa_{m_3} \cap A^\kappa)$ is the lowest priority student who successfully transfers into major $m_3$ at $(\hat{E}^{S2}, \hat{A}^{S2})$. We also have $i_3 \in (\hat{E}^{S2} \cap \hat{A}^{S2}) \subseteq (E' \cap A')$. 
Now, $i_3 \in A'$ implies $i_2 \in A'$; similarly, $i_1 \in E'$ implies $i_2 \in E'$. So $i_2 \in E' \cap A'$ is improved at $(E',A')$. 
However, $m_3 \notin J^k$ and $\tilde{M} \subseteq J^k$ imply $m_3 \notin \tilde{M}$. So $i_2 \in \alpha_{m_3}$ should not be improved at $(E',A')$, which causes a contradiction. 
\qed

\noindent \textbf{Proof of Proposition~\ref{prop_incom2}}: 
Proposition~\ref{prop_incom} has shown that in a major transition program with ceiling and floor constraints, no major transition mechanism can be efficient and have eligibility maximization. 
Therefore, we will prove this by showing the statement still holds for the remaining two situations. 

\underline{Situation (1).} Consider a major transition program with ceiling constraints. We use an example to illustrate it. 
There are three majors $M=\{m_1,m_2,m_3\}$ and three students $I=\{i_1,i_2,i_3\}$. 
    The distributional constraints, transfer-out priority, and transfer-in priority of each major are shown in the following table. 
\begin{center}
\begin{tabular}{|c|c|l|l|l|}
\hline
    Major  & $(\underline{q}_m, \overline{q}_m)$ & transfer-out priority & transfer-in priority \\\hline
    $m_1$  & $(0,1)$ & $\Pout_{m_1}: i_1$ & $\Pin_{m_1}: i_2$ \\\hline
    $m_2$  & $(0,1)$ & $\Pout_{m_2}: i_2$ & $\Pin_{m_2}: i_3, i_1$ \\\hline
    $m_3$  & $(0,1)$ & $\Pout_{m_3}: i_3$ & $-$ \\\hline
\end{tabular}
\end{center}   

Outcome $(I, \emptyset)$ is the unique permissible EM outcome. 
However, it is not Pareto efficient because it can be Pareto dominated by the outcome $(\{i_1,i_2\}, I)$. 
Consider any major transition mechanism that has eligibility maximization. 
In this example, the mechanism must produce the unique permissible EM outcome $(I, \emptyset)$, which is not Pareto efficient. 

\underline{Situation (2).} Consider a major transition program with floor constraints. 
We use an example to illustrate it. 
There are three majors $M=\{m_1,m_2,m_3\}$ and three students $I=\{i_1,i_2,i_3\}$. 
    The distributional constraints, transfer-out priority, and transfer-in priority of each major are shown in the following table. 
\begin{center}
\begin{tabular}{|c|c|l|l|l|}
\hline
    Major  & $(\underline{q}_m, \overline{q}_m)$ & transfer-out priority & transfer-in priority \\\hline
    $m_1$  & $(2,3)$ & $\Pout_{m_1}: i_1, i_2$ & $\Pin_{m_1}: i_3$ \\\hline
    $m_2$  & $(1,3)$ & $\Pout_{m_2}: i_3$ & $\Pin_{m_2}: i_2$ \\\hline
    $m_3$  & $(0,3)$ & $-$ & $\Pin_{m_3}: i_1$ \\\hline
\end{tabular}
\end{center}   

Outcome $(\emptyset, I)$ is the unique permissible EM outcome. 
However, it is not Pareto efficient because it can be Pareto dominated by the outcome $(I, \{i_2, i_3\})$. 
Consider any major transition mechanism that has eligibility maximization. 
In this example, the mechanism must produce the unique permissible EM outcome $(\emptyset, I)$, which is not Pareto efficient. 
\qed



\noindent \textbf{Proof of Proposition~\ref{Prop_diss01}}: 
Consider the processes of the mechanism $f^{E2}$. According to Lemma \ref{Le_dis_01}, every major $m \in M$ has $E_m=\omega_m$. 
Since every major has no transfer-out applicants on its waiting list, no transfer-out exchangeable cycles exist, given $(E, A)$. 
Therefore, the outcome produced by the ToE process in stage $1$ of mechanism $f^{E2}$ is $(\tilde{E}^{S1}, \tilde{A}^{S1})=(E, A)$, so that the outcome produced by the TiE process in stage $2$ of mechanism $f^{E2}$ is $(\tilde{E}^{S2}, \tilde{A}^{S2})$. This is the same as the outcome $(\hat{E}^{S1}, \hat{A}^{S1})$ produced by the TiE process in stage $1$ of mechanism $f^{E1}$. 
That is, $(\tilde{E}^{S2}, \tilde{A}^{S2})=(\hat{E}^{S1}, \hat{A}^{S1})$. 
According to Theorem \ref{Th_maineff2}, the final outcome produced by the mechanism $f^{E2}$ is Pareto efficient, that is, the outcome $(\hat{E}^{S1}, \hat{A}^{S1})$ is Pareto efficient. 

Now consider the processes of the mechanism $f^{E1}$. 
Given the permissible outcome $(E, A)$, the outcome produced by the TiE process in stage $1$ of mechanism $f^{E1}$ is $(\hat{E}^{S1}, \hat{A}^{S1})$. 
Since the outcome $(\hat{E}^{S1}, \hat{A}^{S1})$ is permissible, 
according to Lemma \ref{Le_ToE}, the outcome $(\hat{E}^{S2}, \hat{A}^{S2})$ produced by the ToE process in stage $2$ of mechanism $f^{E1}$ has either $\mu(\hat{E}^{S2}, \hat{A}^{S2})=\mu(\hat{E}^{S1}, \hat{A}^{S1})$ or Pareto dominates the outcome $(\hat{E}^{S1}, \hat{A}^{S1})$. 
We have shown that the outcome $(\hat{E}^{S1}, \hat{A}^{S1})$ is Pareto efficient. 
Therefore, $\mu(\hat{E}^{S2}, \hat{A}^{S2})=\mu(\hat{E}^{S1}, \hat{A}^{S1})$ holds true. 
\qed


\section{Data}\label{App_data}

\begin{table}[H]
\setlength{\abovecaptionskip}{0cm}
\setlength{\belowcaptionskip}{-0.2cm}
\caption{The data for the major transition program operated by a university in China in 2023} \label{Tab_mtdata}
\begin{center}
\footnotesize
\begin{adjustbox}{width=\textwidth}
\begin{tabular}{ccccccc}
\toprule
\textbf{\shortstack{ Major\\ Index}} & \textbf{\shortstack{Transfer-out \\ Applicants}} & \textbf{\shortstack{Actual\\Transfer-out}} & \textbf{\shortstack{Transfer-out\\ Success Rate}} & \textbf{\shortstack{Transfer-in \\Applicants} } & \textbf{\shortstack{Actual \\Transfer-in}} & \textbf{\shortstack{Transfer-in\\ Success Rate}} \\
\midrule
1 & 17 & 9 & 52.94\% & 25 & 7 & 28.00\% \\
2 & 22 & 7 & 31.82\% & 27 & 18 & 66.67\% \\
3 & 23 & 13 & 56.52\% & 22 & 18 & 81.82\% \\
4 & 14 & 8 & 57.14\% & 36 & 27 & 75.00\% \\
5 & 13 & 9 & 69.23\% & 9 & 7 & 77.78\% \\
6 & 138 & 111 & 80.43\% & 189 & 128 & 67.72\% \\
7 & 0 & 0 & - & 162 & 51 & 31.48\% \\
8 & 1 & 1 & 100.00\% & 4 & 4 & 100.00\% \\
9 & 31 & 16 & 51.61\% & 16 & 14 & 87.50\% \\
10 & 34 & 11 & 32.35\% & 34 & 23 & 67.65\% \\
11 & 9 & 6 & 66.67\% & 57 & 28 & 49.12\% \\
12 & 24 & 11 & 45.83\% & 51 & 46 & 90.20\% \\
13 & 30 & 22 & 73.33\% & 5 & 3 & 60.00\% \\
14 & 11 & 8 & 72.73\% & 10 & 9 & 90.00\% \\
15 & 31 & 15 & 48.39\% & 6 & 6 & 100.00\% \\
16 & 102 & 64 & 62.75\% & 1 & 1 & 100.00\% \\
17 & 66 & 41 & 61.19\% & 9 & 8 & 88.89\% \\
18 & 16 & 13 & 81.25\% & 20 & 18 & 90.00\% \\
19 & 80 & 52 & 65.00\% & 1 & 1 & 100.00\% \\
20 & 76 & 44 & 57.89\% & 0 & 0 & - \\
21 & 54 & 30 & 55.56\% & 51 & 42 & 82.35\% \\
22 & 8 & 4 & 50.00\% & 80 & 30 & 37.04\% \\
23 & 50 & 32 & 64.00\% & 5 & 5 & 100.00\% \\
24 & 30 & 14 & 46.67\% & 11 & 9 & 81.82\% \\
25 & 9 & 5 & 55.56\% & 0 & 0 & - \\
26 & 6 & 4 & 75.00\% & 79 & 45 & 56.96\% \\
27 & 5 & 2 & 40.00\% & 0 & 0 & - \\
28 & 24 & 9 & 37.50\% & 1 & 1 & 100.00\% \\
29 & 0 & 0 & - & 22 & 12 & 54.55\% \\
30 & 55 & 31 & 56.36\% & 40 & 30 & 75.00\% \\
31 & 12 & 7 & 58.33\% & 6 & 5 & 83.33\% \\
32 & 25 & 16 & 64.00\% & 39 & 21 & 53.85\% \\
\midrule
\textbf{Total} & 1018 & 617 & 60.55\% & 1018 & 617 & 60.55\% \\
\bottomrule
\end{tabular}
\end{adjustbox}
\end{center}
\end{table}

\section{Mechanism}\label{App_mech}

We say that a major $m$ is \textbf{transfer-out expandable} at an outcome $(E, A)$ if (1) $\bar{E}_m \neq \emptyset$;
(2) either $\max_{\Pout_m} \bar{E}_m \notin A$ or $\vert \mu_m (E, A) \vert > \underline{q}_m$. 
The condition (1) indicates that not every $m$'s transfer-out applicants obtain the transfer-out eligibility from $m$. 
The condition (2) means that major $m$ can add an additional transfer-out eligibility to the top priority student among students from $\bar{E}_m$ without violating its dual priority and floor constraint. 
Let $M^{\mathrm{out+}}(E, A)$ be the set of transfer-out expandable majors at outcome $(E, A)$, 
and 
$M^{\mathrm{in-}}(E, A)=\{m \in M \mid \vert \mu_m(E, A) \vert > \overline{q}_m \}$ be the set of majors whose floor constraint is violated at $(E, A)$. 
A formal description of the alternative EM mechanism is as follows.

\begin{center}
    \textbf{An alternative mechanism with eligibility maximization}
\end{center}
\begin{itemize}    
    \item[\textbf{Step $0$}] Let $E^1 = \emptyset$, $A^1 = I$. 
    \item[\textbf{Step $k$}]($k \geq 1$) Consider outcome $(E^k, A^k)$. Do the following processes: 
    \begin{itemize}
        \item[$k.1$]  (\textit{Transfer-out process}) 
        If $M^{\mathrm{out+}}(E^k, A^k)=\emptyset$, go to step k.2. 
        Otherwise, each major $m \in M^{\mathrm{out+}}(E^k, A^k)$ distributes transfer-out eligibility to the student $\max_{\Pout_m}\bar{E}^k_m$ who has the highest priority among those transfer-out applicants who have not yet obtained transfer-out eligibility from $m$. 
        Set $E^{k+1}=E^k \cup (\max_{\Pout_m} \bar{E}^k_m)_{m \in M^{\mathrm{out+}}(E^k, A^k)}$ and $A^{k+1}=A^k$. Let $k=k+1$, and go to step k.

        \item[$k.2$]  (\textit{Transfer-in process}) 
        If $M^{\mathrm{in-}}(E^k, A^k)=\emptyset$, stop. 
        Otherwise, each major $m \in M^{\mathrm{in-}}(E^k, A^k)$ revokes the transfer-in eligibility of the student $\min_{\Pin_m}A^k_m$ who has the lowest priority among those transfer-in applicants who have obtained transfer-in eligibility from $m$. 
        Set $A^{k+1}= A^k \setminus (\min_{\Pin_m}A^k_m)_{m \in M^{\mathrm{in-}}(E^k, A^k)}$ and $E^{k+1}=E^k$. Let $k=k+1$, and go to step k.

    \end{itemize}    
\end{itemize}

\begin{proposition}
    The alternative EM mechanism produces a permissible EM outcome for every problem.
\end{proposition}

Since the alternative EM mechanism and the EM mechanism are symmetric, we omit the proof.

\section{Simulations}\label{App_simulation}

In this section, we run simulations under various scenarios to compare the performances of the CMT-EC mechanism, the EM mechanism, and the two efficient major transition mechanisms. 
We consider an environment with 10 majors, in which each major initially has 100 students. 
Each student is allocated to an initial major. 
We construct the preference profile of each student $i \in I$ by taking the possible correlation among students' preferences into account. 
Specifically, we calculate student $i$'s utility from being assigned to a major $m$ as follows,
\[
U(i,m)=\beta_1A(m)+\beta_2B(m)+(1-\beta_1-\beta_2)C(\omega_i)
\]
where $A(m) \in (0,1)$ is an i.i.d. standard uniformly distributed random variable, which represents the individual taste of $i$ on $m$; $B(m) \in (0,1)$ is an i.i.d. standard uniformly distributed random variable, which  represents the common tastes of students on $m$; $C(\omega_i)=1$ represents the utilities for the students' initial majors. 
The correlation in the students' preferences is based on $\beta_1, \beta_2 \in [0,1]$. 
When $\beta_1$ increases, the students' preferences over majors become more similar. 
When $\beta_2$ increases, the students' preferences over majors become more individualistic. 
When $(1-\beta_1-\beta_2)$ increases, students attribute more value to their initial majors. 
Each student $i$ applies to the major that has the maximum utility, i.e., $\max_{m \in M} U(i,m)$.\footnote{\ If student $i$ has $\max_{m \in M} U(i,m)=\omega_i$, the student will not apply to a new major. }

For each major $m$, we draw values from an i.i.d. standard uniform distribution for transfer-out students to determine their transfer-out priority. Similarly, we draw values from an i.i.d. standard uniform distribution for transfer-in students to determine their transfer-in priority for that major. 

We simulate two types of major transition programs. First, we consider the major transition program under the balanced condition, such that for each major $m \in M$, its floor constraint and ceiling constraint are equal to the current capacity, i.e., $\underline{q}_m=\overline{q}_m=100$. 
Second, we consider the major transition program in which the balanced condition is not necessary. 
In particular, we set each major $m$'s floor constraint and ceiling constraint as $90\%$ and $110\%$ of its current capacity, respectively, that is $\underline{q}_m=90$ and $\overline{q}_m=110$. 

We simulate various scenarios for each type of major transition program. 
Specifically, these scenarios can be divided into three groups. 
In the first group, we set the parameter $(1-\beta_1-\beta_2)=0.4$, and the parameter $\beta_1$ ranges from $0.3$ to $0.6$ and $\beta_2$ ranges from $0$ to $0.3$. 
In the second group, we set the parameter $(1-\beta_1-\beta_2)=0.3$, and the parameter $\beta_1$ ranges from $0.4$ to $0.7$ and $\beta_2$ ranges from $0$ to $0.3$. 
In the third group, we set the parameter $(1-\beta_1-\beta_2)=0.2$, and the parameter $\beta_1$ ranges from $0.4$ to $0.7$ and $\beta_2$ ranges from $0.1$ to $0.4$. 

In each scenario, we run the CMT-EC mechanism, the EM mechanism, and the two efficient mechanisms, $f^{E1}$ and $f^{E2}$, 1,000 times for each considered correlation of students' preferences, and calculate the successful transition rate (STR) under these mechanisms, respectively, where 
\[
STR=\frac{\text{the number of students who successfully transfer their majors}}{\text{the number of students who want to apply to a new major}}.
\]

\subsection{Major Transition Program with Balanced Condition}

\begin{figure}[ht]  
    \centering
    \includegraphics[width=\textwidth]{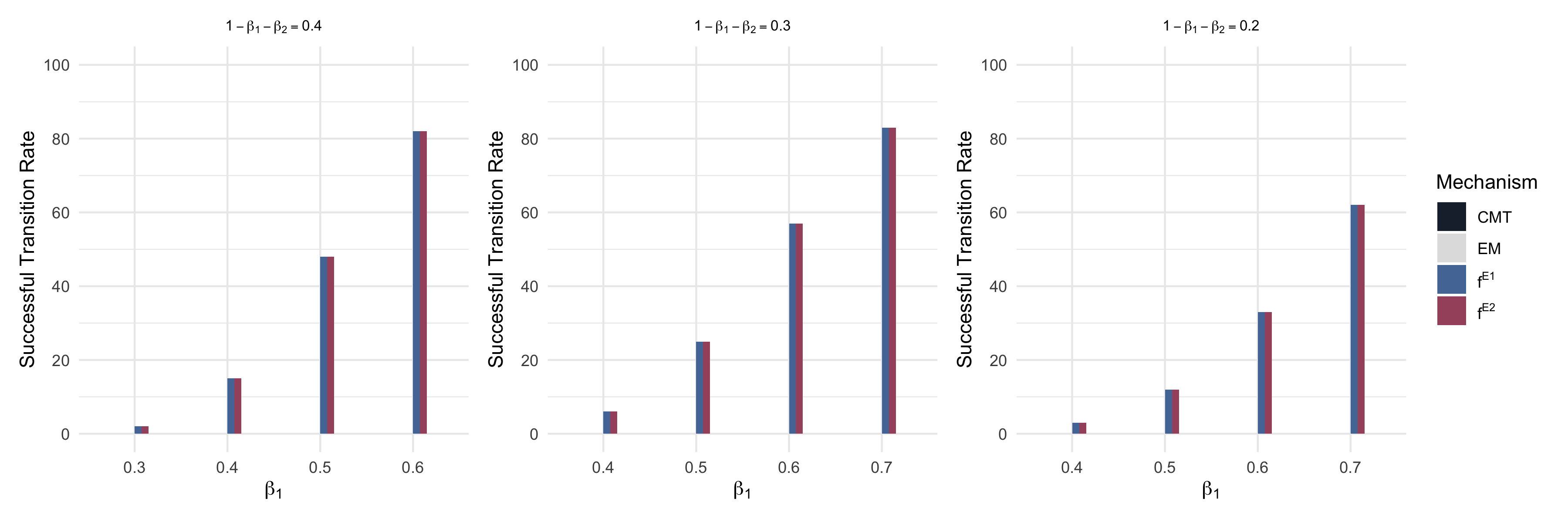}
    \caption{Successful Transition Rate (STR) in Program with Balanced Condition}
    \label{fig:simulation_balanced}  
\end{figure}

Figure \ref{fig:simulation_balanced} illustrates the successful transition rate under the CMT-EC mechanism, the EM mechanism, and the two efficient mechanisms $f^{E1}, f^{E2}$ under each scenario. 
We have the following main findings. 
\begin{itemize}
\item[(a)] The efficiency of the CMT-EC mechanism and the EM mechanism is zero, which implies that the two mechanisms fail to work when a balanced condition is required. 
In other words, in this case, the two efficient mechanisms $f^{E1}, f^{E2}$ play important roles in finding Pareto efficient outcomes. 

\item[(b)] The STR of the two efficient mechanisms $f^{E1}, f^{E2}$ are the same under each scenario. 

\item[(c)] When the parameter $(1-\beta_1-\beta_2)$ is fixed, with the increase of $\beta_1$, the STR of the efficient mechanisms increases. For instance, in the case that $(1-\beta_1-\beta_2)=0.4$, the STR of the efficient mechanisms is $1.8\%$ when $\beta_1=0.3$, while STR of the efficient mechanisms increases to $81.7\%$ when $\beta_1=0.6$. 
This implies that when students' preferences are more individualistic, they are more likely to transfer major successfully. 

\item[(d)] When the parameter $\beta_1$ is fixed, with the decreases of $(1-\beta_1-\beta_2)$, the number of students who apply to a new major increases and the STR of the efficient mechanisms decreases. For instance, in the case that $\beta_1=0.5$, the STR-E is $48.3\%$ when $1-\beta_1-\beta_2=0.4$, the STR of the efficient mechanisms decreases to $24.65\%$ when $1-\beta_1-\beta_2=0.3$, and further decreases to $11.93\%$ when $1-\beta_1-\beta_2=0.2$. 
This implies that when students value their majors less, more students want to transfer their majors, which increases the competition among applicants, leading to a lower STR of the efficient mechanisms. 

\end{itemize}

\subsection{Major Transition Program without Balanced Condition}

\begin{figure}[ht]  
    \centering
    \includegraphics[width=\textwidth]{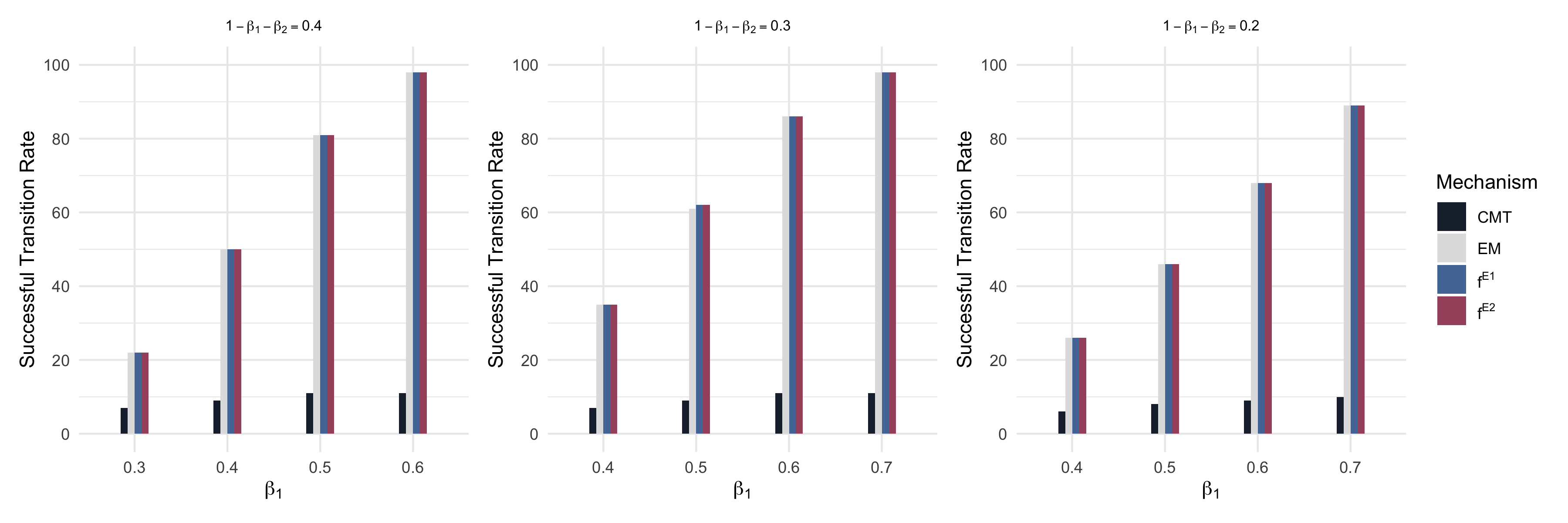}
    \caption{Successful Transition Rate (STR) in Program without Balanced Condition}
    \label{fig:simulation_wo_balanecd}  
\end{figure}

Figure \ref{fig:simulation_wo_balanecd} illustrates the successful transition rate under the CMT-EC mechanism, the EM mechanism, and the two efficient mechanisms, $f^{E1}$ and $f^{E2}$, under each scenario. 
We have the following main findings. 

\begin{itemize} 

   \item[(e)] The efficiency of the CMT-EC mechanism is significantly lower than that of the EM and the two efficient mechanisms, $f^{E1}$ and $f^{E2}$. For example, when $ (1-\beta_1-\beta_2) = 0.4$  and $\beta_1 = 0.6$, the STR of the CMT-EC mechanism is $11.65\%$, whereas the STR of the EM mechanism and the two efficient mechanisms is $98.36\%$.

    \item[(f)] The efficiency of the two efficient mechanisms is the same and the efficiency of the EM is almost the same as that of the two efficient mechanisms under each scenario. 
    This implies that when no balanced condition is required in the major transition program, the EM mechanism performs almost as well as the two efficient mechanisms.

    \item[(g)] Similar to finding (c), when the parameter $(1-\beta_1-\beta_2)$ is fixed, with the increase of $\beta_1$, the STR-EM increases. For instance, in the case that $(1-\beta_1-\beta_2)=0.4$, the STR of the EM mechanism is $57.45\%$ and the STR-E is $57.52\%$ when $\beta_1=0.3$, while the STR of the EM mechanism is $98.35\%$ and the STR-E is $98.36\%$ when $\beta_1=0.6$
    
    \item[(h)] Similar to finding (d), when the parameter $\beta_1$ is fixed, with the decreases of $(1-\beta_1-\beta_2)$, the number of students who apply to a new major increases and the STR-E decreases.

\end{itemize}

To sum up, the CMT-EC mechanism performs worst in both situations, with and without the balanced condition. 
In general, the two efficient mechanisms $f^{E1}, f^{E2}$ perform the best, and the more individualistic are students' preferences, the higher is the successful transition rate of the efficient mechanisms. 
It is worth mentioning that the EM mechanism shows significant differences in performance. 
When the balanced condition is required, the performance of the EM mechanism is as poor as the CMT mechanism, in which the successful transition rate is zero. 
However, when the balanced condition is not required, the EM mechanism surprisingly performs as well as the two efficient mechanisms, such that the successful transition rate is almost the same as those of the efficient mechanisms under every scenario.

\end{document}